\documentclass[12pt]{article}
\usepackage{amsmath,amssymb,amsthm,amsxtra,overpic,bm,epsfig}
\usepackage{color}
\usepackage{comment}
\usepackage[all,cmtip]{xy}
\usepackage{multirow}
\usepackage{hyperref}
\usepackage{graphicx}
\graphicspath{{figures/}}
\usepackage{dcolumn}
\usepackage{url}            
\usepackage{booktabs}       
\usepackage{amsfonts}       
\usepackage{nicefrac}       
\usepackage{microtype}      
\usepackage{lipsum}
\usepackage{mathrsfs}
\usepackage{indentfirst}
\usepackage{subcaption}
\usepackage{float}
\usepackage[utf8]{inputenc}
\usepackage{cite}
\usepackage{slashed,stmaryrd}
\usepackage{subcaption}
\renewcommand{\thefootnote}{\fnsymbol{footnote}}

\begin{document}

\vspace{0.2cm}

\begin{center}
{\large\bf Search for dark photon in association with charged leptons at electron-positron colliders}
\end{center}

\vspace{0.2cm}

\begin{center}
{\bf Dai-Xing Zhang}\textsuperscript{a} \footnote{E-mail:24412011130@stumail.sdut.edu.cn}, \quad
{\bf Chang-Qi Hu}\textsuperscript{b} \footnote{E-mail:hcq116029@163.com}, \quad
{\bf Chun-Yuan Li}\textsuperscript{a} \footnote{E-mail:lichunyuan@sdut.edu.cn}, \quad 
{\bf Xiao Liang}\textsuperscript{a} \footnote{E-mail:lx321@sdut.edu.cn}, \quad 
{\bf Xing-Hua Yang}\textsuperscript{a} \footnote{E-mail:yangxinghua@sdut.edu.cn} \quad 


\textsuperscript{a}{School of Physics and Optoelectronic Engineering, Shandong University of Technology, Zibo, Shandong 255000, China}\\
\textsuperscript{b}{Department of Physics, Liaoning Normal University,  Dalian, Liaoning 116029, China}\\

\end{center}

\vspace{1.5cm}

\begin{abstract}

The dark photon can serve as a vector portal between the dark sector and the Standard Model via kinetic mixing.
In this paper, we investigate the production of an invisible dark photon via the process $e^+ e^- \to \ell^+ \ell^- A'$ with $\ell=e,\mu$, at the Circular Electron Positron Collider (CEPC).
The production cross sections for the signal process are calculated, and the kinematic distributions for the signal and background processes are analyzed.
We further compute the branching ratio for the invisible decay of the dark photon, and explore the discovery potential of dark photons at the CEPC.
It is found that dark photons with masses in the range $1~\mathrm{GeV} < M_{A'} < 16~\mathrm{GeV}$ are potentially detectable at the CEPC.

\end{abstract}

\begin{flushleft}
\hspace{0.88cm} {\bf Keywords}:  dark photon, CEPC, invisible decay
\end{flushleft}

\def\thefootnote{\arabic{footnote}}
\setcounter{footnote}{0}

\newpage

\section{Introduction}\label{sec1}

Despite the remarkable success of the Standard Model (SM) in describing the properties and interactions of elementary particles, it is widely recognized that the SM is not a complete theory of particle physics.
In particular, the microscopic nature of dark matter and the mechanisms mediating its interactions with SM particles remain open problems. 
These open questions have stimulated extensive investigations of dark sector models.
Such models connect the dark sector to the SM through the introduction of new particles and new interactions~\cite{Jaeckel:2010ni,Essig:2013lka,Lanfranchi:2020crw}.
Among these models, the vector portal mediated by a dark photon is one of the simplest scenarios.
The dark photon is the massive gauge boson of the $U(1)_D$ symmetry, and it can interact with SM particles through kinetic mixing with the electromagnetic field~\cite{Holdom:1985ag,Fabbrichesi:2020wbt}.
The dark photon can decay into SM particles, producing visible final states, and can also decay into dark matter particles, producing invisible final states.
Its decay modes are determined by the masses of the dark sector particles and their couplings to the dark photon.

Searches for dark photons have been performed at various experimental facilities, including fixed-target and beam-dump experiments~\cite{Bjorken:1988as,E141,E774,APEX:2011dww,Battaglieri:2014hga,Balewski:2014pxa,NA64_2019,NA64_2023,LDMX:2018cma,PADME:2022xly} and collider experiments~\cite{BaBar,BaBar:2017tiz,Belle,KLOE-2:2015nli,BESIII:2025otp,LHCb:2017,LHCb:2019vmc,ATLAS:2019itm,ATLAS:2022xlo,ATLAS:2023jyp,CMS:2019buh,CMS:2020krr}.
These searches cover broad regions of the dark photon mass and kinetic mixing parameter space.
Astronomical and cosmological observations have also placed certain constraints on the dark photon parameter space~\cite{Fradette:2014sza,An:2022hhb,Chang:2016ntp,Arsenadze:2024ywr}.
Ref.~\cite{Caputo:2026pdw} summarizes the current constraints on the dark photon parameter space in the $(M_{A'}, \epsilon)$ plane from various experimental searches. 
In the GeV mass range, the most stringent constraints are currently imposed by BaBar, CMS, and LHCb.
The BaBar experiment found no significant excess over the SM background in the dark photon mass range of $0.02$--$10.2~\mathrm{GeV}$ and set $90\%$ CL upper limits on the kinetic mixing parameter $\epsilon$, with the limits lying in the range $10^{-4}$--$10^{-3}$~\cite{BaBar}.
The CMS Collaboration searched for a narrow dimuon resonance in the mass range $11.5$--$200~\mathrm{GeV}$, observed no significant excess, and set the most stringent limits in the mass ranges $30$--$75~\mathrm{GeV}$ and $110$--$200~\mathrm{GeV}$~\cite{CMS:2019buh}.
LHCb searched for prompt-like dark photons with masses up to $70~\mathrm{GeV}$ and for long-lived dark photons with masses in the $214$--$350~\mathrm{MeV}$ range, but found no evidence of a signal.
In the mass range $10.6 < M_{A'} < 70~\mathrm{GeV}$, exclusion limits on the kinetic mixing parameter are set at $90\%$ CL~\cite{LHCb:2017}.
Although collider experiments have placed relatively stringent constraints on dark photons with masses ranging from $1~\mathrm{GeV}$ to several hundred $\mathrm{GeV}$, substantial regions of the parameter space remain unexplored and warrant further investigation.

Many theorists have also studied dark photons through various processes at different colliders~\cite{Curtin:2014cca,He:2017zzr,He:2017ord,Berggren:2026qbl,Jiang:2019hfn,Inan:2021dir,Liang:2026rqp,BESIII:2024rkp,Li:2025tlg,Chen:2025hbh,Gao:2024rgl,Casarsa:2021rud}.
Ref.~\cite{Curtin:2014cca} studied dark photon searches at hadron colliders through Drell--Yan production, $pp \to A' \to \ell^+\ell^-$, and exotic Higgs decays.
Ref.~\cite{Berggren:2026qbl} studied dark photon searches in visible decay channels at electron--positron colliders. The search is based on the radiative process $e^+e^-\to \gamma A'\to \gamma \ell^+\ell^-$, in which the dilepton resonance can be used to identify the dark photon.
For invisible decays, Ref.~\cite{Inan:2021dir} studied recoil-based searches via the process $\gamma e^-\to e^-A'$ at the ILC and CLIC.
In Ref.~\cite{Liang:2026rqp}, the authors searched for light dark photons through quarkonium decays, while Ref.~\cite{BESIII:2024rkp} studied rare charm decays.
In addition, dark final state radiation $A'\to A' s$ at future electron--positron colliders was studied in Ref.~\cite{Li:2025tlg}, while axion--photon--dark photon interactions were studied in Ref.~\cite{Chen:2025hbh}.
Ref.~\cite{Gao:2024rgl} studied coherent electron bremsstrahlung at electron--ion colliders and showed that it provides sensitivity in the dark photon mass range $2m_e$--$3~\mathrm{GeV}$.
Future muon colliders can extend the reach to heavier dark photons via $\mu^+\mu^-\rightarrow\gamma A'$~\cite{Casarsa:2021rud}.
In this work, we study dark photon production via $e^+e^- \to \ell^+\ell^- A'$ at the CEPC, considering the center-of-mass energies $\sqrt{s}=91.2~\mathrm{GeV}$ and $240~\mathrm{GeV}$ and dark photon masses in the range $1$--$200~\mathrm{GeV}$.
The kinematic distributions between the signal and background processes are discussed, and the invisible decays of the dark photon and their branching ratios are taken into account. 
We further evaluate the potential of the CEPC to probe dark photons.

The rest of the paper is organized as follows.
In Sec.~\ref{sec:theory}, the theoretical framework of the dark photon model is presented.
In Sec.~\ref{sec:signal}, we calculate the signal cross sections and study the signal and background kinematic distributions.
In Sec.~\ref{sec:result}, the event selection is described and the significance is calculated.
Sec.~\ref{sec:conclusion} concludes with a summary.

\section{\label{sec:theory}Theoretical Framework of the Dark Photon}

\begin{figure}[htbp]
    \centering
    \includegraphics[width=0.95\textwidth]{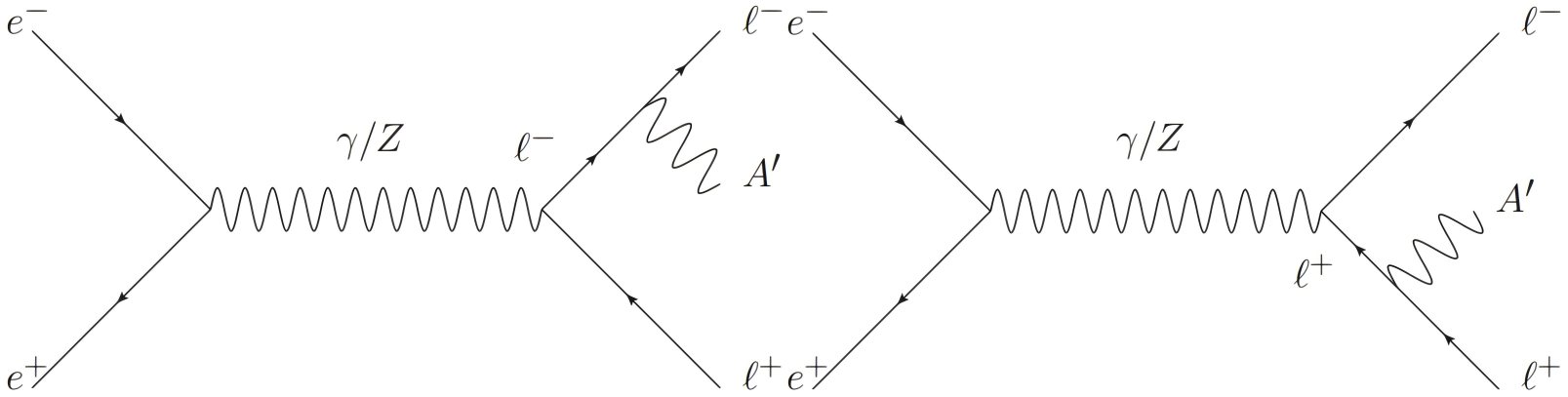}
    \caption{Feynman diagrams for the signal process $e^+e^-\rightarrow\ell^+\ell^-A'$.}
    \label{fig:1a}
\end{figure}

Renormalizable portals provide a natural framework for connecting the dark sector to the SM~\cite{Agrawal:2021dbo,Battaglieri:2017aum,Alexander:2016aln}.
Among these portals, we focus on the vector portal. This portal introduces a new Abelian gauge field, which can kinetically mix with the SM photon field; its gauge boson is the dark photon.
Kinetic mixing induces an effective parity conserving coupling to the electromagnetic current, with an effective coupling strength of $\epsilon e$, which is suppressed relative to the elementary charge $e$ by the factor $\epsilon$~\cite{Essig:2013lka,Bjorken:2009mm}.
After diagonalization of the kinetic mixing, the relevant Lagrangian of the dark photon model is:
\begin{equation}
\mathcal{L}_{\rm eff}
\supset
-\frac{1}{4}F'_{\mu\nu}F'^{\mu\nu}
+\frac{1}{2}M_{A'}^2A'_\mu A'^\mu
-\epsilon eA'_\mu J^\mu_{\rm EM},
\label{eq:full_lagrangian}
\end{equation}
where $A'_\mu$ denotes the dark photon field, $F'_{\mu\nu}$ is the corresponding field strength tensor, and $J^\mu_{\rm EM}$ denotes the electromagnetic current of the SM.
The electromagnetic current takes the form~\cite{Fabbrichesi:2020wbt}:
\begin{equation}
J^\mu_{\rm EM}
=
\sum_{\ell} Q_{\ell}\bar{\ell}\gamma^\mu{\ell} ,
\label{eq:4}
\end{equation}
where $Q_\ell$ denotes the electric charge of the charged lepton $\ell$, expressed in units of the elementary charge $e$.
The effective electromagnetic-current interaction is used throughout this work, while additional contributions associated with electroweak mixing are neglected. 
This approximation should be considered when interpreting the results at high dark photon masses.
In this work, we consider the final-state radiation process.
The signal process is given by
\begin{equation}
e^+e^-\rightarrow\ell^+\ell^-A' ,
\end{equation}
where the charged leptons in the final state are restricted to the electron and muon channels, namely $\ell=e,\mu$. 
The dark photon is emitted from the charged lepton lines via the kinetic mixing, as illustrated by the Feynman diagrams in Fig.~\ref{fig:1a}.

We also calculate the total decay width and the branching ratio for the invisible decay.
The total decay width of the dark photon is obtained by summing the visible and invisible decay widths:
\begin{equation}
\Gamma_{\rm tot}
=
\Gamma_{\rm vis}
+
\Gamma_{\rm inv},
\end{equation}
where the visible decays include the leptonic and hadronic decay channels:
\begin{equation}
\Gamma_{\rm vis}
=
\Gamma_{A'\rightarrow e^+e^-}
+
\Gamma_{A'\rightarrow\mu^+\mu^-}
+
\Gamma_{A'\rightarrow{\rm hadrons}} .
\end{equation}
The hadronic contribution to the dark photon decay width is included through
\begin{equation}
\Gamma_{A'\rightarrow {\rm hadrons}}
=
\Gamma_{A'\rightarrow\mu^+\mu^-}R,
\end{equation}
where $R$ denotes the experimentally measured ratio of the hadronic to muonic cross sections in $e^+e^-$ annihilation. 
The hadronic contribution is calculated using the experimentally measured $R$ ratio given by the Particle Data Group~\cite{ParticleDataGroup:2026mpi}.

The decay widths of the dark photon into leptons and dark matter particles are given by~\cite{Fabbrichesi:2020wbt}
\begin{equation}
\begin{aligned}
\Gamma_{A'\rightarrow \ell^+\ell^-}
&=
\frac{\alpha_{\rm em}\epsilon^2 M_{A'}}{3}
\left(1+\frac{2m_\ell^2}{M_{A'}^2}\right)
\sqrt{1-\frac{4m_\ell^2}{M_{A'}^2}},
\\[6pt]
\Gamma_{A'\rightarrow\chi\bar{\chi}}
&=
\frac{\alpha_D M_{A'}}{3}
\left(1+\frac{2m_\chi^2}{M_{A'}^2}\right)
\sqrt{1-\frac{4m_\chi^2}{M_{A'}^2}},
\end{aligned}
\label{eq:decaywidth}
\end{equation}
where $M_{A'}$, $m_\ell$, and $m_\chi$ denote the masses of the dark photon, charged leptons, and dark matter particles. 
The parameters $\epsilon$ and $\alpha_{\rm em}$ represent the kinetic mixing parameter and the electromagnetic fine structure constant, while $\alpha_D$ describes the effective interaction strength between the dark photon and the dark sector.
The branching ratio of the invisible decay is then calculated as follows:
\begin{equation}
\mathrm{Br}(A'\rightarrow\chi\bar{\chi})
=
\frac{\Gamma_{\rm inv}}
{\Gamma_{\rm tot}} .
\label{eq:br}
\end{equation}

\section{\label{sec:signal}Dark Photon Production and Kinematic Distributions}

In this section, dark photon masses in the range $1$--$200~\mathrm{GeV}$ are considered. The cross sections for the signal process are calculated at $\sqrt{s}=91.2~\mathrm{GeV}$ and $240~\mathrm{GeV}$, and the kinematic distributions between the signal and background processes are analyzed.

\begin{figure}[H]
    \centering
    \begin{subfigure}[b]{0.45\textwidth}
        \centering
        \includegraphics[width=\textwidth]{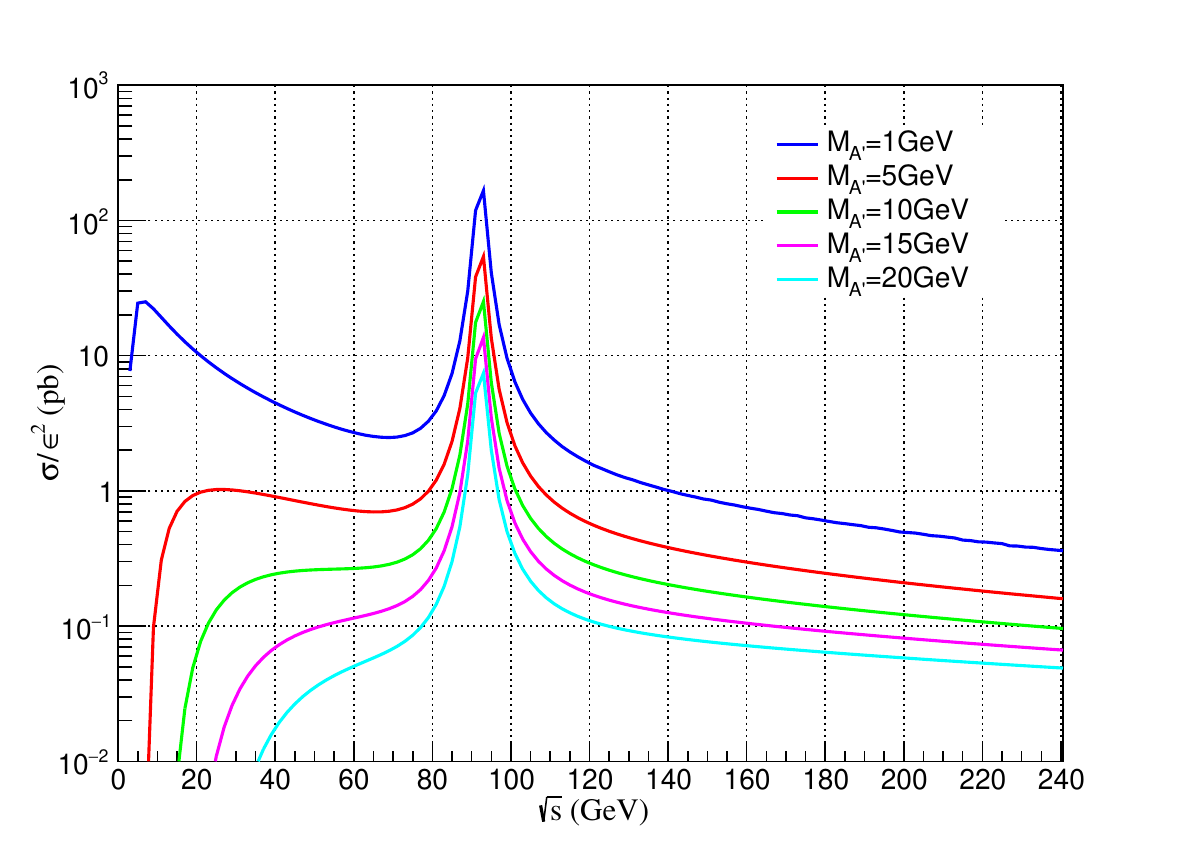}
        \caption{(a)}
        \label{fig:left}
    \end{subfigure}
    \hfill
    \begin{subfigure}[b]{0.45\textwidth}
        \centering
        \includegraphics[width=\textwidth]{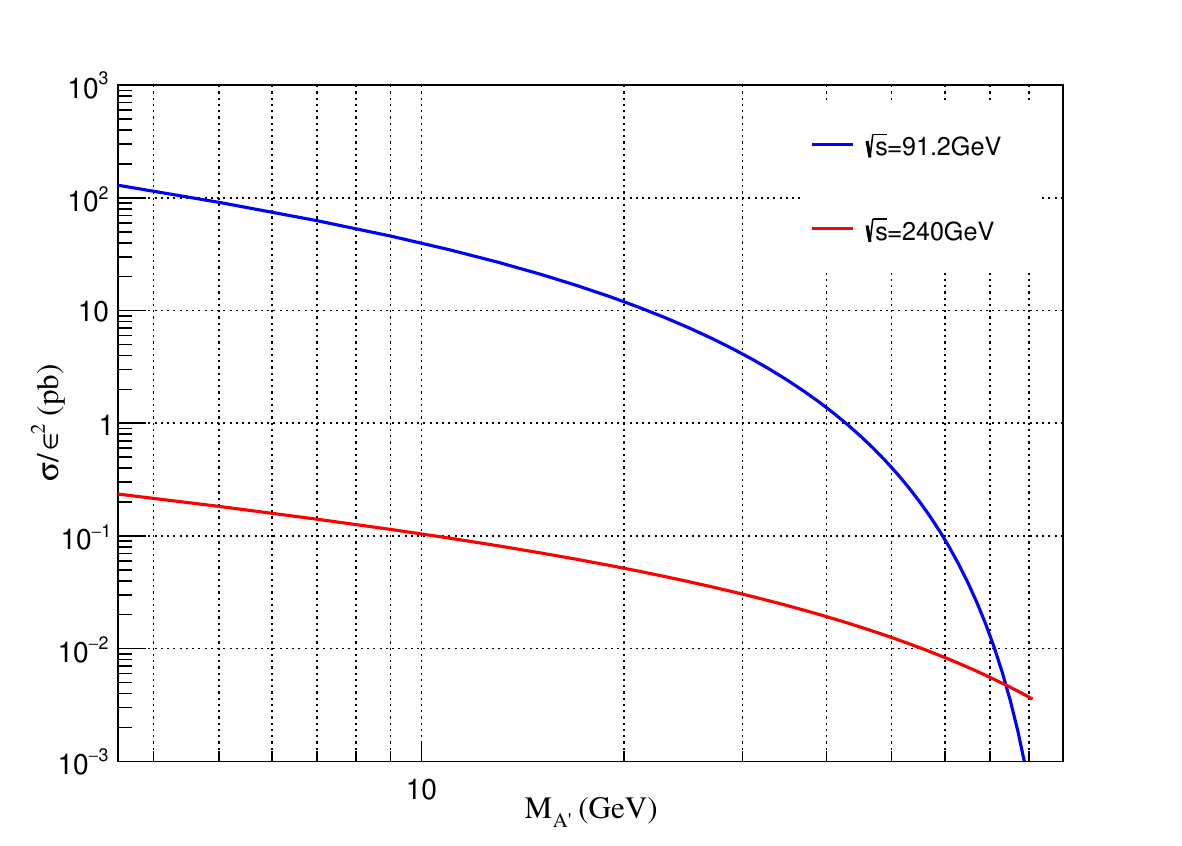}
        \caption{(b)}
        \label{fig:center}
    \end{subfigure}
    \caption{Reduced cross sections $\sigma/\epsilon^2$ for $e^+e^-\to\ell^+\ell^-A'$. (a): dependence on $\sqrt{s}$; (b): dependence on $M_{A'}$ at $\sqrt{s}=91.2$ and $240~\mathrm{GeV}$.}
    \label{fig:2a}
\end{figure}

FeynArts~\cite{Feynarts} and FeynCalc~\cite{FeynCalc} are employed to calculate the analytic amplitudes of the signal process, and the \texttt{Cuba} package~\cite{CuBa} is used to numerically integrate the resulting scattering amplitudes over phase space.
Figure~\ref{fig:2a} shows the reduced cross section for the process $e^+e^- \to \ell^+\ell^- A'$.
The left panel shows $\sigma/\epsilon^2$ as a function of the center-of-mass energy for different dark photon masses; the right panel shows $\sigma/\epsilon^2$ as a function of the dark photon mass for center-of-mass energies $\sqrt{s}=91.2~\mathrm{GeV}$ and $240~\mathrm{GeV}$.
For a given dark photon mass, the reduced cross section exhibits a peak when $\sqrt{s}=M_Z$.
At $\sqrt{s}=91.2~\mathrm{GeV}$, the reduced cross section decreases more rapidly than that at $\sqrt{s}=240~\mathrm{GeV}$; the two curves intersect at $M_{A'} \sim 75~\mathrm{GeV}$.

\begin{figure}[H]
    \centering
    \begin{minipage}{0.3\textwidth}
        \centering
        \includegraphics[width=\linewidth]{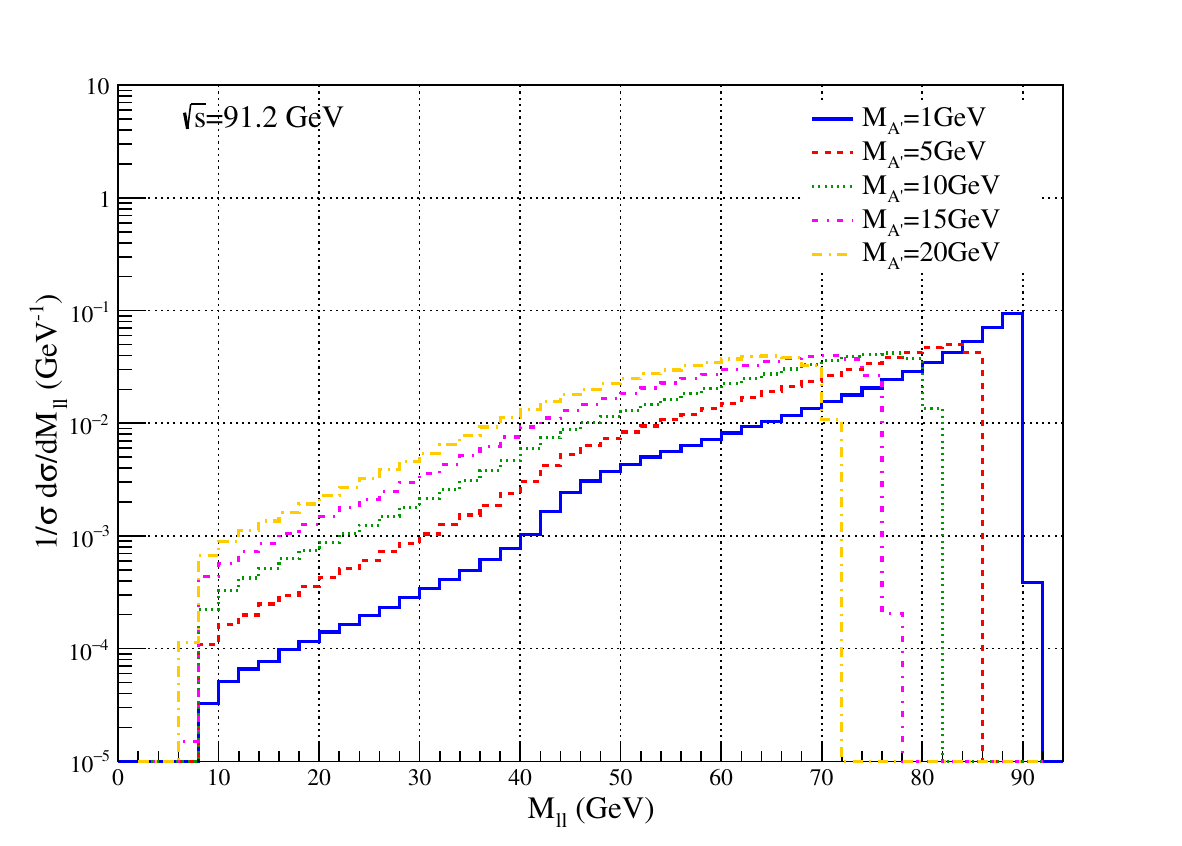}
            \vspace{0.1em}
    {\small (a)}
    \end{minipage}
    \hfill
    \begin{minipage}{0.3\textwidth}
        \centering
        \includegraphics[width=\linewidth]{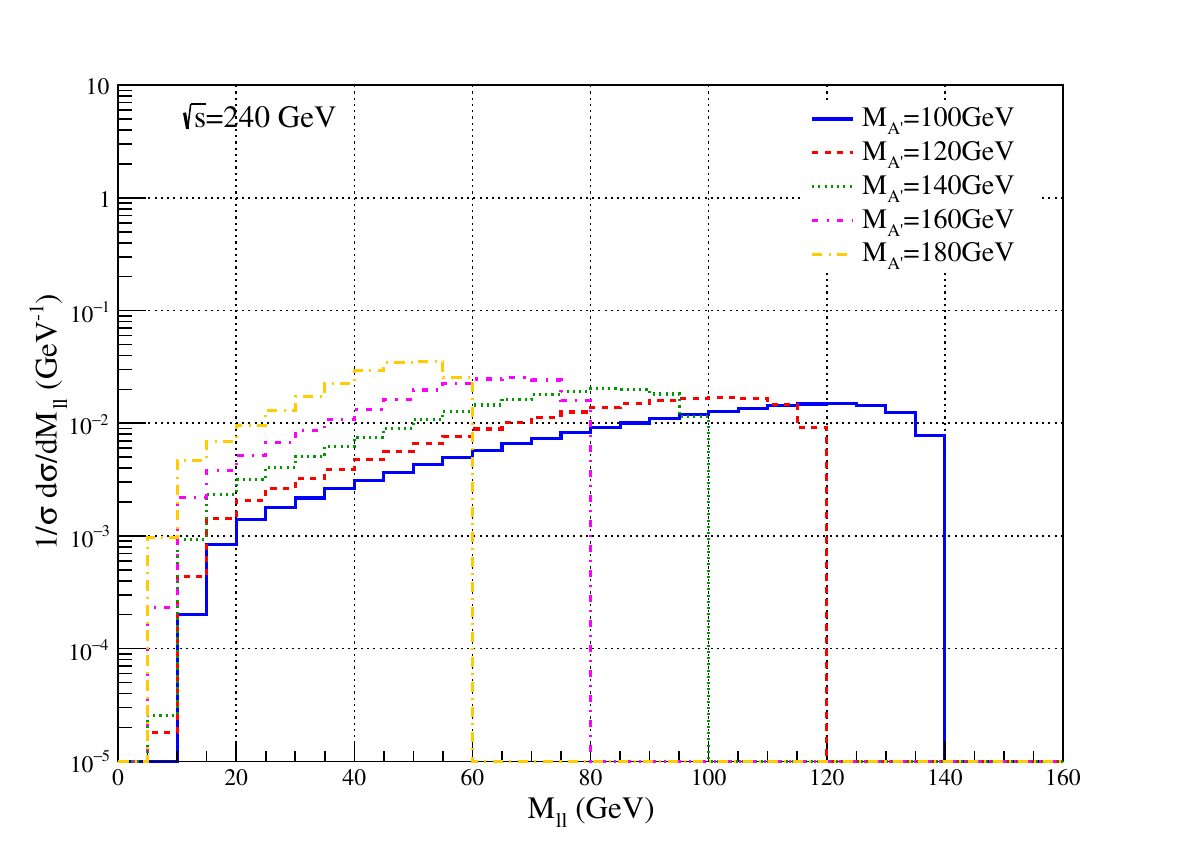}
            \vspace{0.1em}
    {\small (b)}
    \end{minipage}
    \hfill
    \begin{minipage}{0.3\textwidth}
        \centering
        \includegraphics[width=\linewidth]{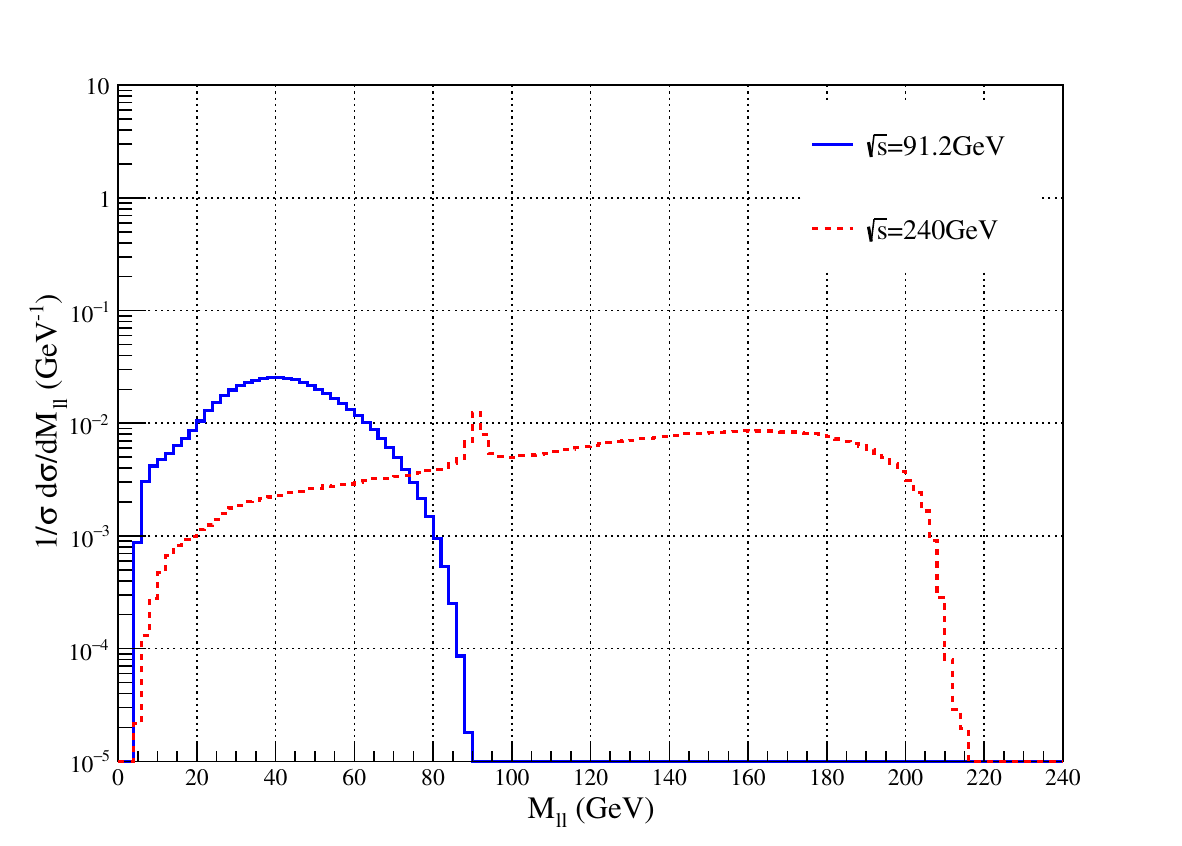}
            \vspace{0.1em}
    {\small (c)}
    \end{minipage}
    \vspace{\baselineskip} 
    \begin{minipage}{0.3\textwidth}
        \centering
        \includegraphics[width=\linewidth]{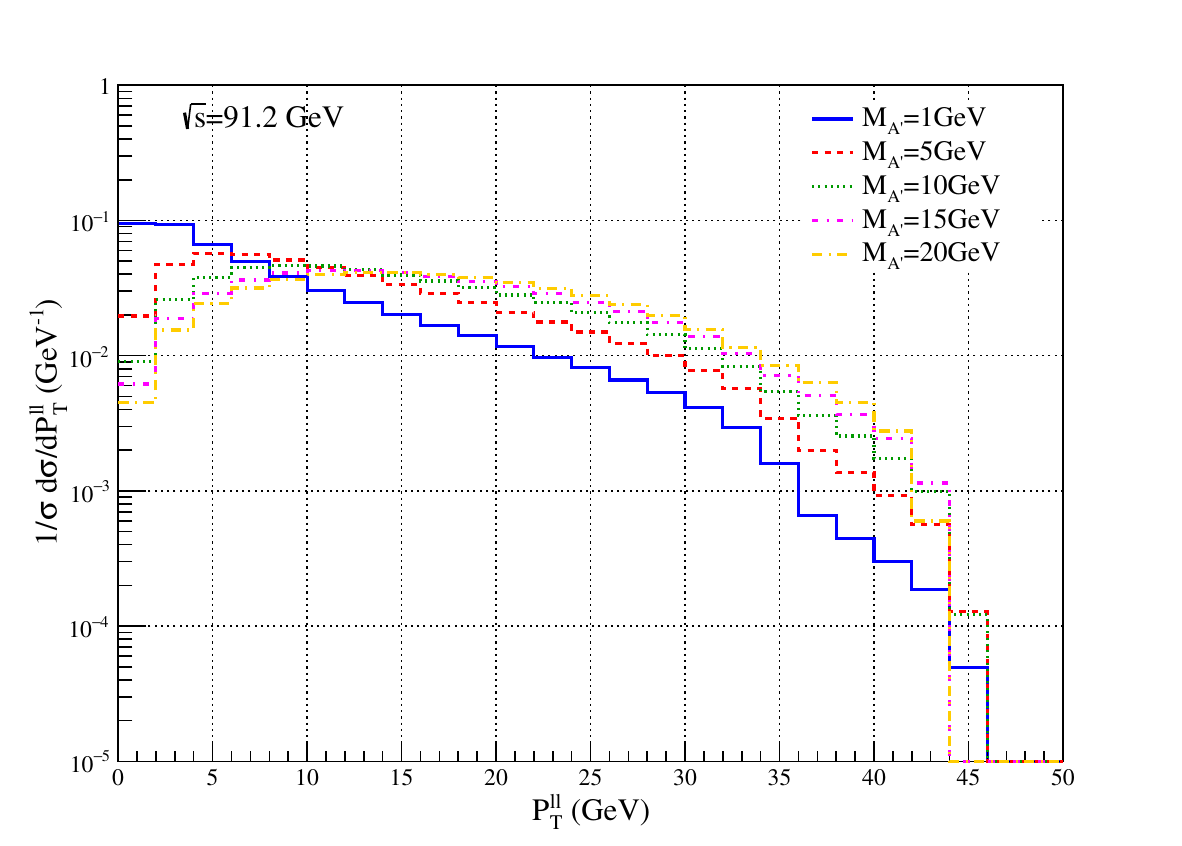}
            \vspace{0.1em}
    {\small (d)}
    \end{minipage}
    \hfill
    \begin{minipage}{0.3\textwidth}
        \centering
        \includegraphics[width=\linewidth]{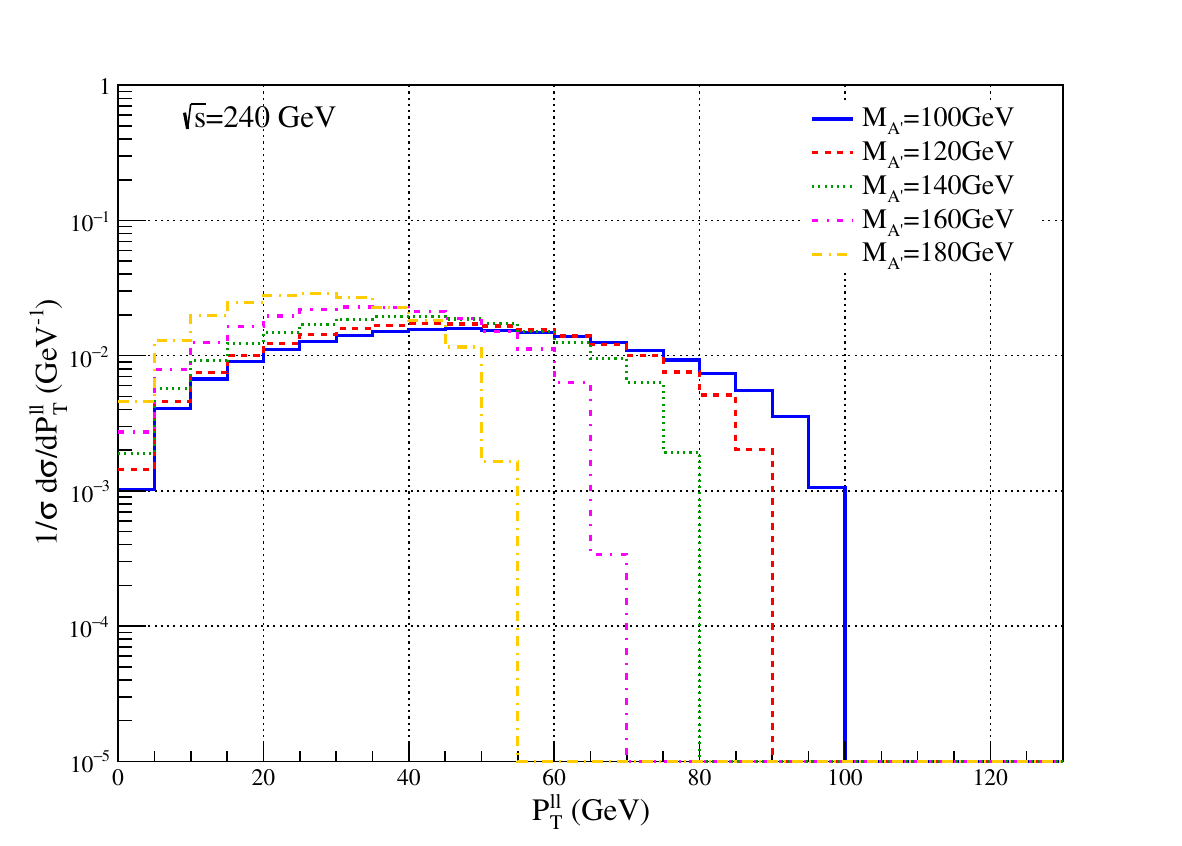}
            \vspace{0.1em}
    {\small (e)}
    \end{minipage}
    \hfill
    \begin{minipage}{0.3\textwidth}
        \centering
        \includegraphics[width=\linewidth]{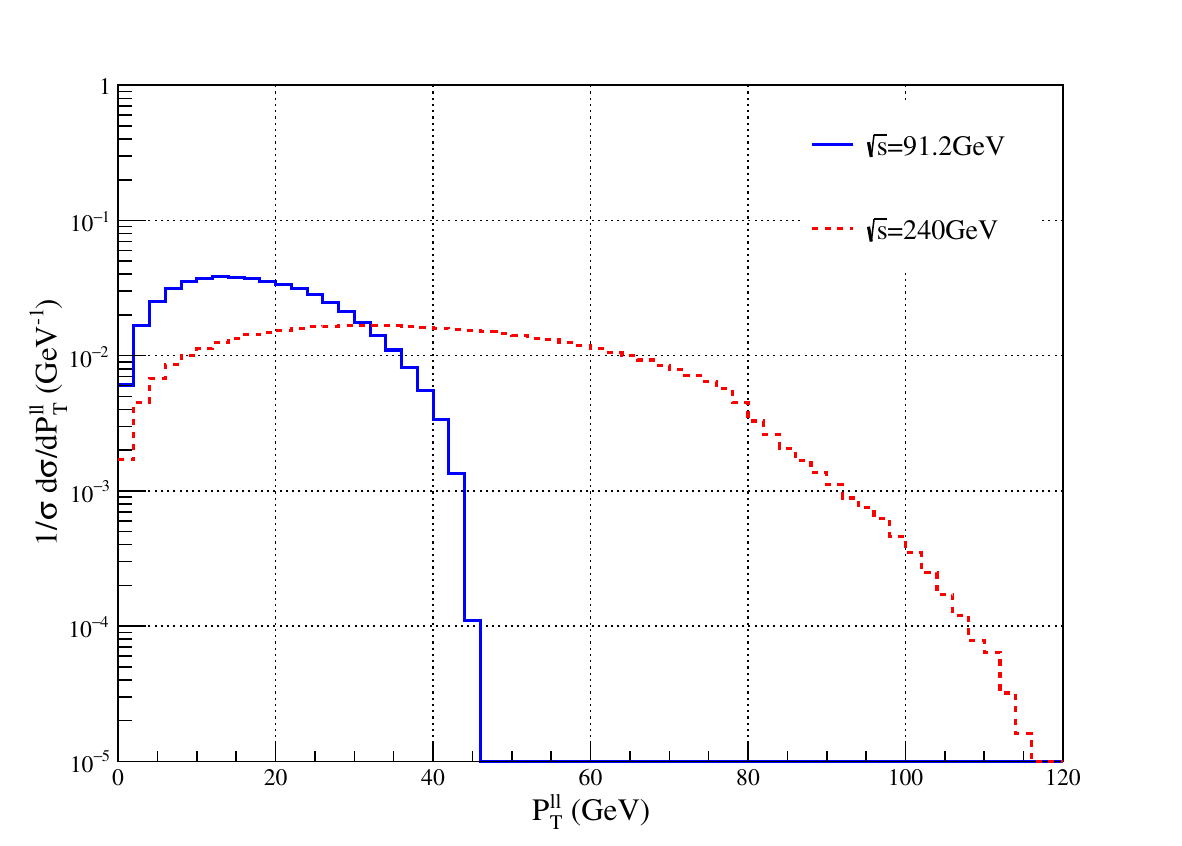}
            \vspace{0.1em}            
    {\small (f)}
    \end{minipage}
    \vspace{\baselineskip}
    \begin{minipage}{0.3\textwidth}
        \centering
        \includegraphics[width=\linewidth]{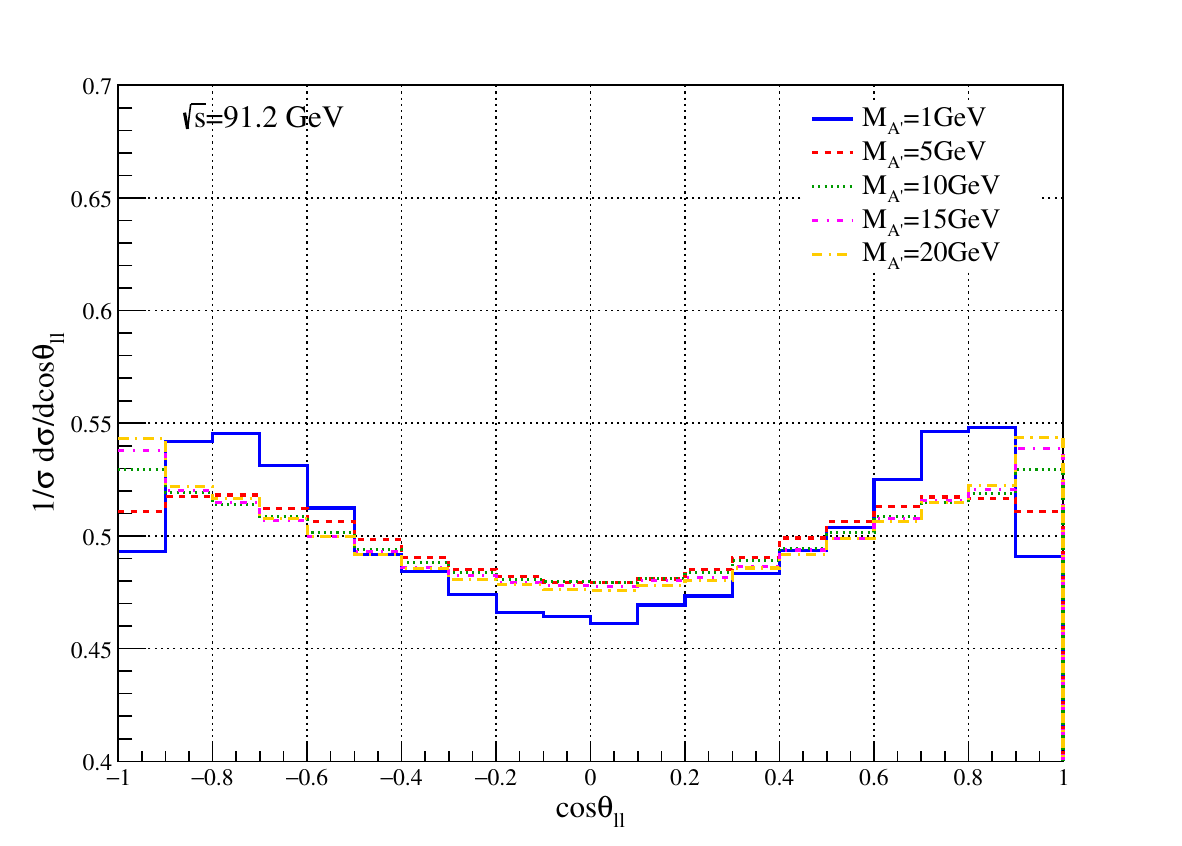}
            \vspace{0.1em}
    {\small (g)}
    \end{minipage}
    \hfill
    \begin{minipage}{0.3\textwidth}
        \centering
        \includegraphics[width=\linewidth]{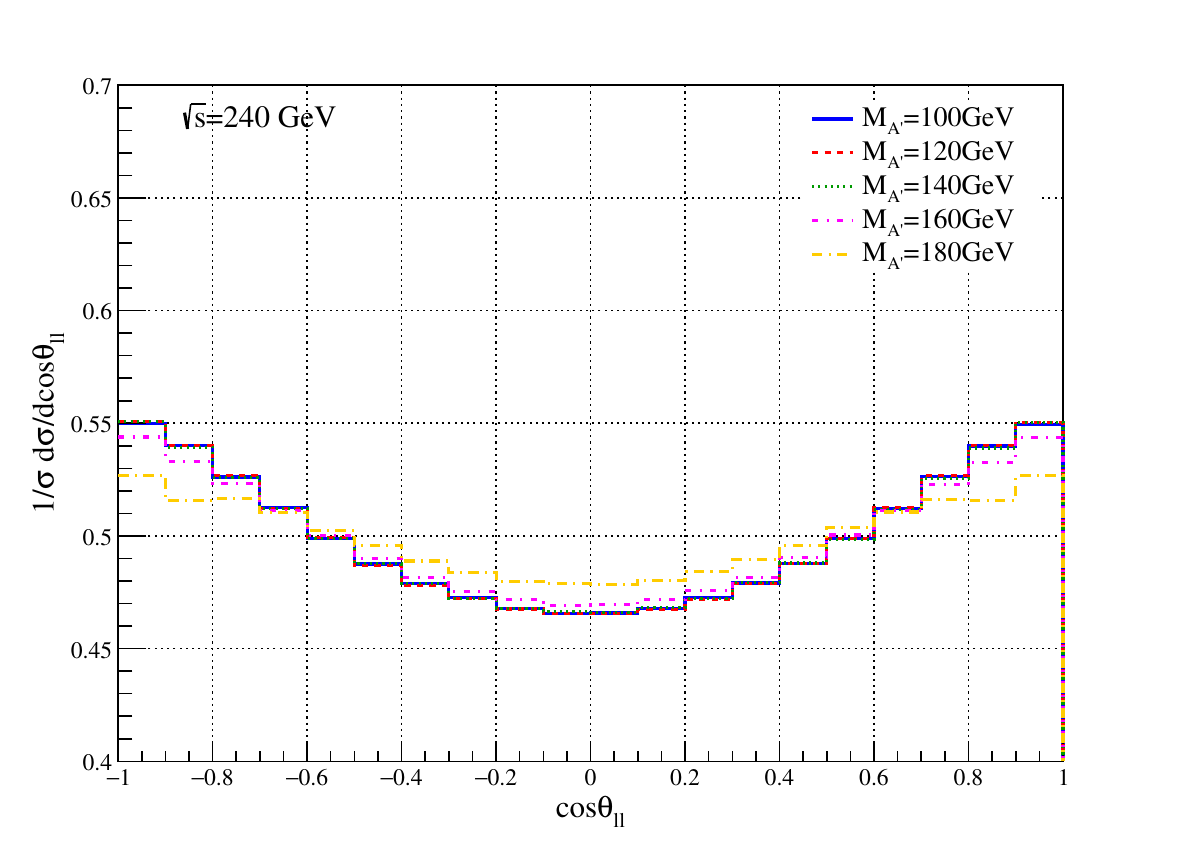}
            \vspace{0.1em}
    {\small (h)}
    \end{minipage}
    \hfill
    \begin{minipage}{0.3\textwidth}
        \centering
        \includegraphics[width=\linewidth]{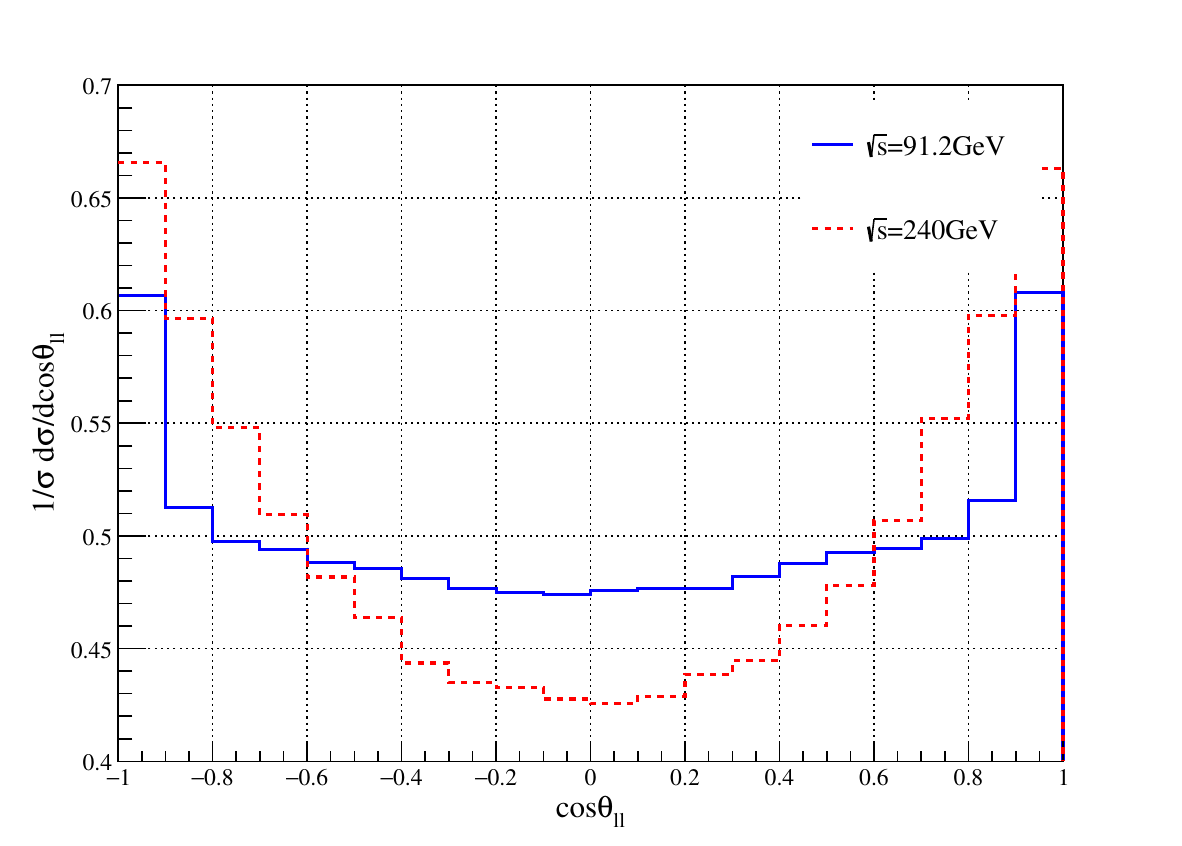}
            \vspace{0.1em}
    {\small (i)}
    \end{minipage}
    \vspace{\baselineskip}
    \begin{minipage}{0.3\textwidth}
        \centering
        \includegraphics[width=\linewidth]{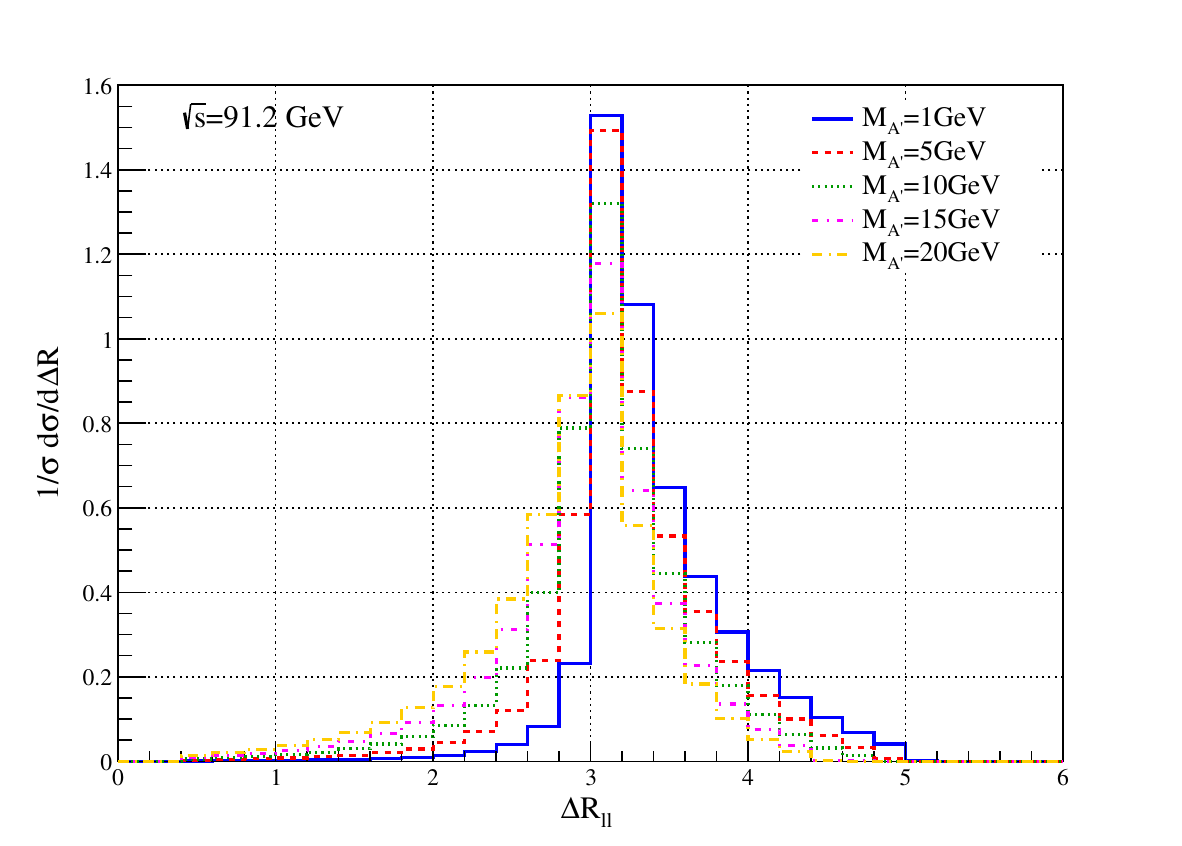}
            \vspace{0.1em}
    {\small (j)}
    \end{minipage}
    \hfill
    \begin{minipage}{0.3\textwidth}
        \centering
        \includegraphics[width=\linewidth]{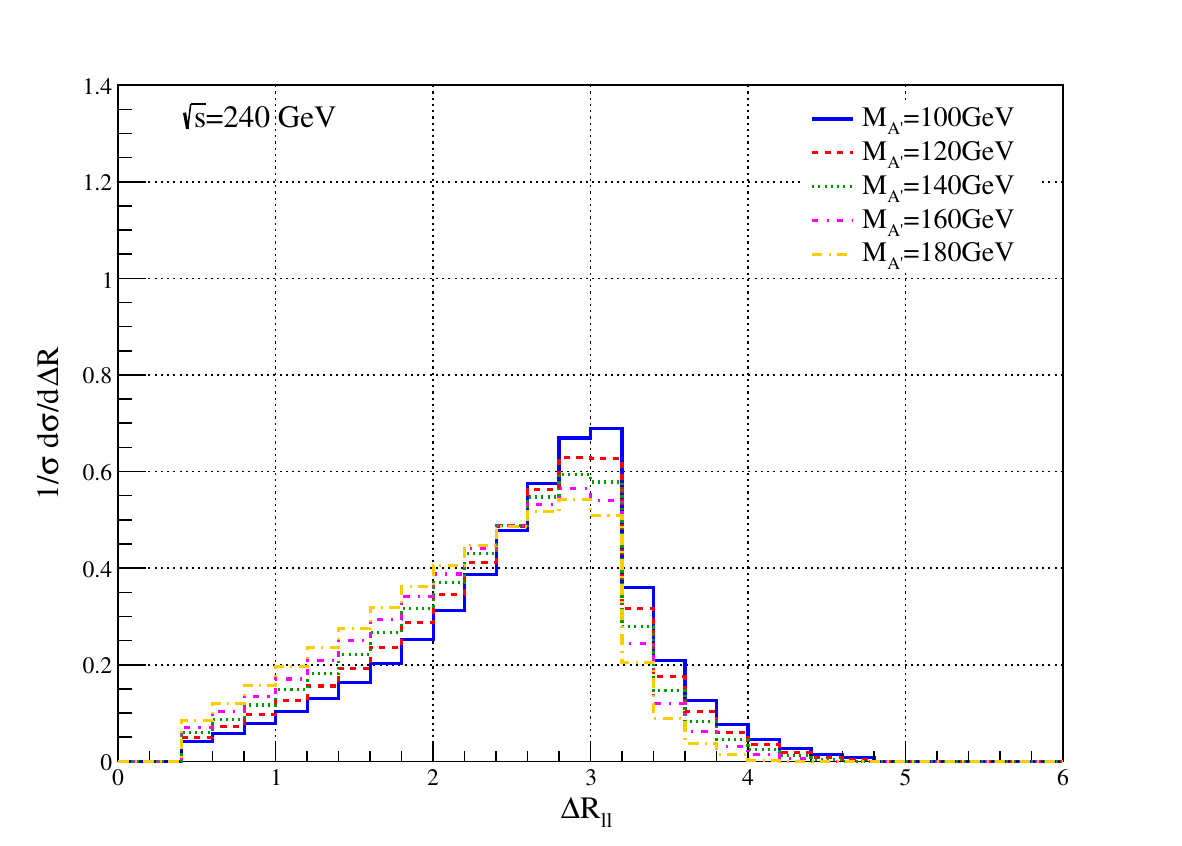}
            \vspace{0.1em}
    {\small (k)}
    \end{minipage}
    \hfill
    \begin{minipage}{0.31\textwidth}
        \centering
        \includegraphics[width=\linewidth]{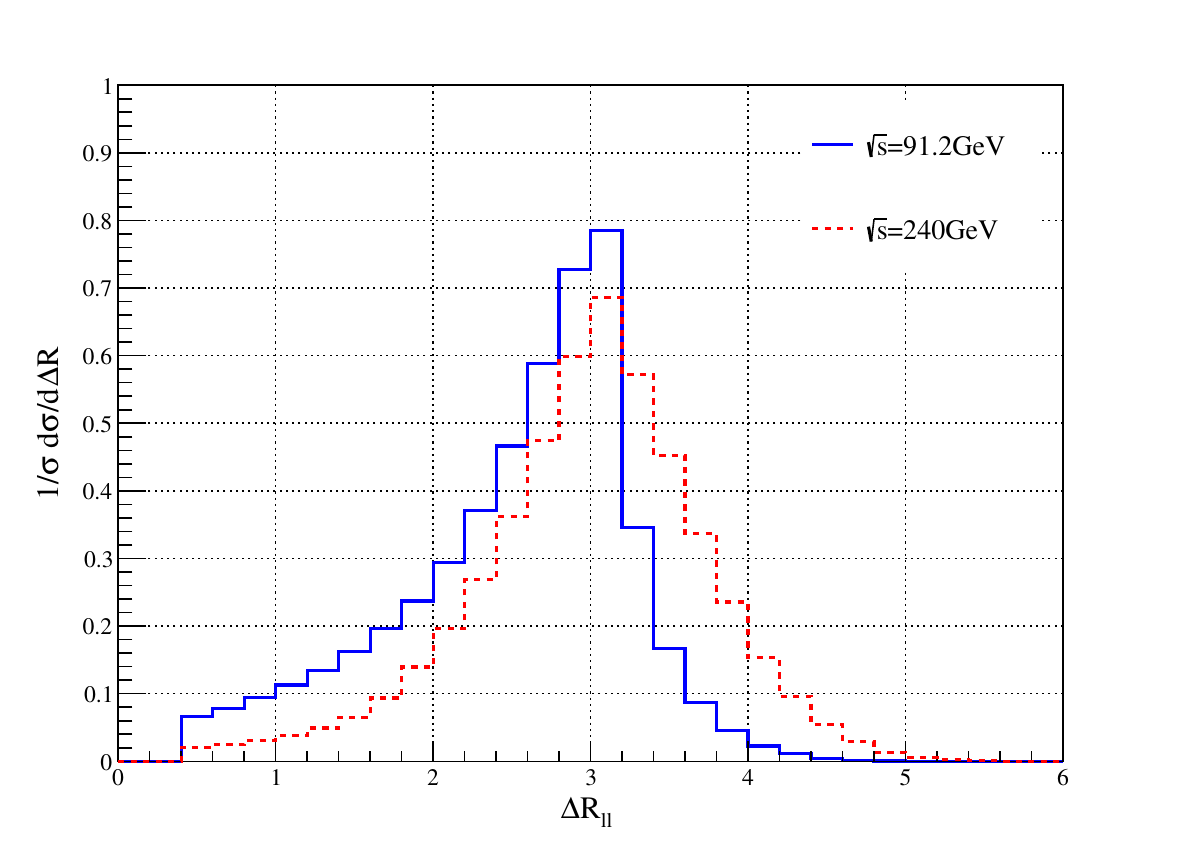}
            \vspace{0.1em}
    {\small (l)}
    \end{minipage}   
\caption{
Normalized kinematic distributions of $M_{\ell\ell}$, $p_T^{\ell\ell}$, $\cos\theta_{\ell\ell}$, and $\Delta R_{\ell\ell}$ for the signal and the SM background processes at $\sqrt{s}=91.2~\mathrm{GeV}$ and $240~\mathrm{GeV}$.
}
    \label{fig:3a}
\end{figure}

We further investigate the kinematic distributions of the signal process and the corresponding SM background processes.
The process $e^+e^- \to \ell^+\ell^- \nu_\ell \bar{\nu}_\ell$ is generated as the background process using \texttt{MadGraph5\_aMC@NLO}~\cite{Alwall:2014hca}.
Based on the CEPC studies in Refs.~\cite{CEPCStudyGroup:2018ghi,CEPCPhysicsStudyGroup:2022uwl,Gao:2022lew}, we adopt the following basic selection criteria for signal and background events in our numerical analysis
\begin{equation}
p_T^\ell>10~\mathrm{GeV},\qquad |\eta_\ell|<2.5,\qquad \Delta R_{\ell\ell}>0.4 .
\end{equation}
After applying the above selection criteria, the kinematic distributions of the signal and background processes at the CEPC for $\sqrt{s}=91.2$ and $240~\mathrm{GeV}$ are shown in Fig.~\ref{fig:3a}.
The kinematic variables in the figures are defined as follows.
The dilepton invariant mass is defined as $M_{\ell\ell}=\sqrt{(p_{\ell^+}+p_{\ell^-})^2}$, where $p_{\ell^\pm}$ are the four-momenta of the charged leptons.
The transverse momentum of the dilepton system is defined as $p_T^{\ell\ell}=\sqrt{(p_x^{\ell\ell})^2+(p_y^{\ell\ell})^2}$, where $p^{\ell\ell}=p_{\ell^+}+p_{\ell^-}=(E_{\ell\ell},\,p_x^{\ell\ell},\,p_y^{\ell\ell},\,p_z^{\ell\ell})$ is the four-momentum of the dilepton system.
The cosine of the polar angle is defined as $\cos\theta_{\ell\ell} = p_z^{\ell\ell} \big/ \sqrt{(p_x^{\ell\ell})^2 + (p_y^{\ell\ell})^2 + (p_z^{\ell\ell})^2}$.
The angular separation between the two charged leptons is defined as $\Delta R_{\ell\ell} = \sqrt{(\Delta\eta)^2 + (\Delta\phi)^2}$, where $\Delta\eta$ and $\Delta\phi$ are the differences in pseudorapidity and azimuthal angle, respectively, with $\Delta\phi$ taken in the range $[-\pi,\pi]$.

We investigate the detection of dark photon at the CEPC over the mass range $1$--$200\mathrm{GeV}$ using two center-of-mass energies: at $\sqrt{s}=91.2\mathrm{GeV}$, the considered dark photon mass range is $1$--$20\mathrm{GeV}$, while at $\sqrt{s}=240\mathrm{GeV}$, it is $20$--$200\mathrm{GeV}$.
To illustrate the specific kinematic distributions, we select several representative dark photon mass points: $M_{A'}=1$, $5$, $10$, $15$, and $20~\mathrm{GeV}$ at $\sqrt{s}=91.2~\mathrm{GeV}$, and $M_{A'}=100$, $120$, $140$, $160$, and $180~\mathrm{GeV}$ at $\sqrt{s}=240~\mathrm{GeV}$, and present the corresponding distributions.

As shown in Fig.~\ref{fig:3a}, at $\sqrt{s}=91.2~\mathrm{GeV}$, the normalized kinematic distribution increases with $M_{\ell\ell}$ and then drops rapidly near the upper endpoint of the $M_{\ell\ell}$ distribution, while the SM background distribution first increases and then decreases with increasing $M_{\ell\ell}$, peaking around $40~\mathrm{GeV}$. 
At $\sqrt{s}=240~\mathrm{GeV}$, the signal distribution becomes broader and flatter, and the SM background distribution shows a distinct peak at $M_{\ell\ell}=M_Z$ due to the resonance of the $Z$ boson.
For the signal process, the allowed upper endpoint of the $M_{\ell\ell}$ distribution decreases as the dark photon mass increases.
At $\sqrt{s}=91.2~\mathrm{GeV}$, for $p_T^{\ell\ell}>10~\mathrm{GeV}$, the normalized $p_T^{\ell\ell}$ distribution decreases with increasing $p_T^{\ell\ell}$.
Both the signal and background distributions rise slowly and then decrease at $\sqrt{s}=240~\mathrm{GeV}$. 
The upper endpoint of the $p_T^{\ell\ell}$ distribution is also related to the dark photon mass $M_{A'}$.
The normalized $\cos\theta_{\ell\ell}$ distribution exhibits a concave trend, and near $|\cos\theta_{\ell\ell}|=1$ there is some difference between the signal and background distributions.
For the $\Delta R_{\ell\ell}$ distribution at $\sqrt{s}=91.2~\mathrm{GeV}$, the signal distribution is concentrated mainly in the region $\Delta R_{\ell\ell}>3$, whereas the background distribution is concentrated mainly in the region $\Delta R_{\ell\ell}<3$. 
Although the signal distribution becomes broader as $M_{A'}$ increases, this separation remains evident.
At $\sqrt{s}=240~\mathrm{GeV}$, the signal distribution is instead concentrated mainly in the region $\Delta R_{\ell\ell}<3$, with a less pronounced peak.

\begin{figure}[H]
    \centering
    \begin{subfigure}{0.44\textwidth}
        \centering
        \includegraphics[width=\linewidth]{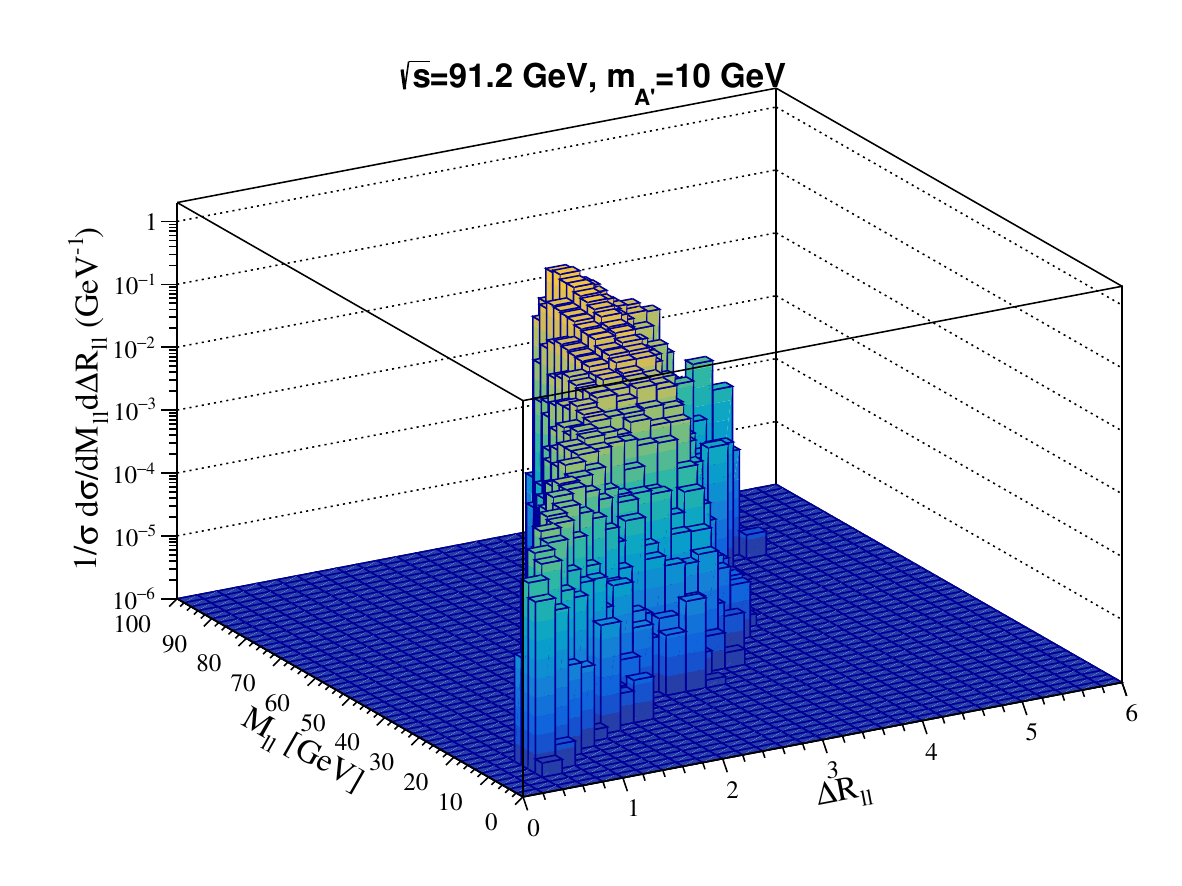}
        \caption{(a)}
    \end{subfigure}
    \hfill
    \begin{subfigure}{0.44\textwidth}
        \centering
        \includegraphics[width=\linewidth]{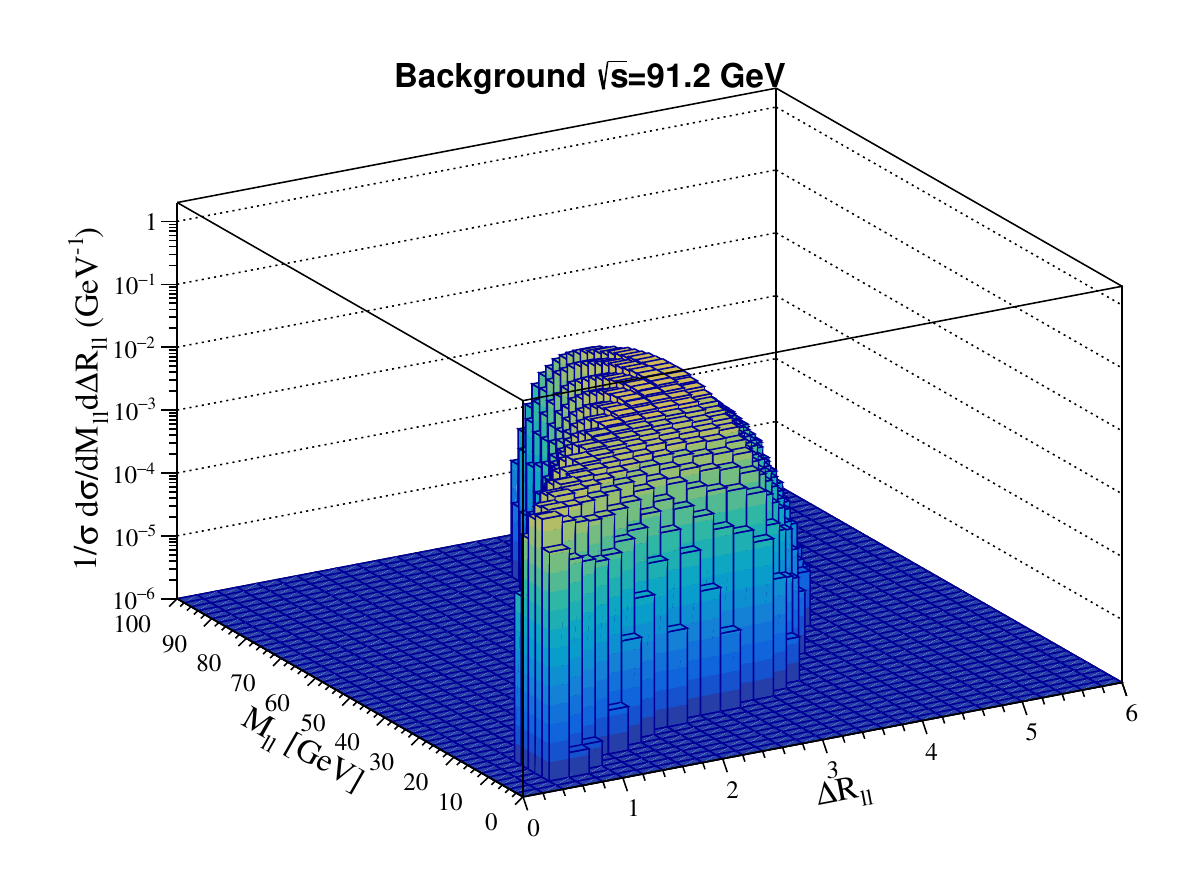}
        \caption{(b)}
    \end{subfigure}
    \begin{subfigure}{0.44\textwidth}
        \centering
        \includegraphics[width=\linewidth]{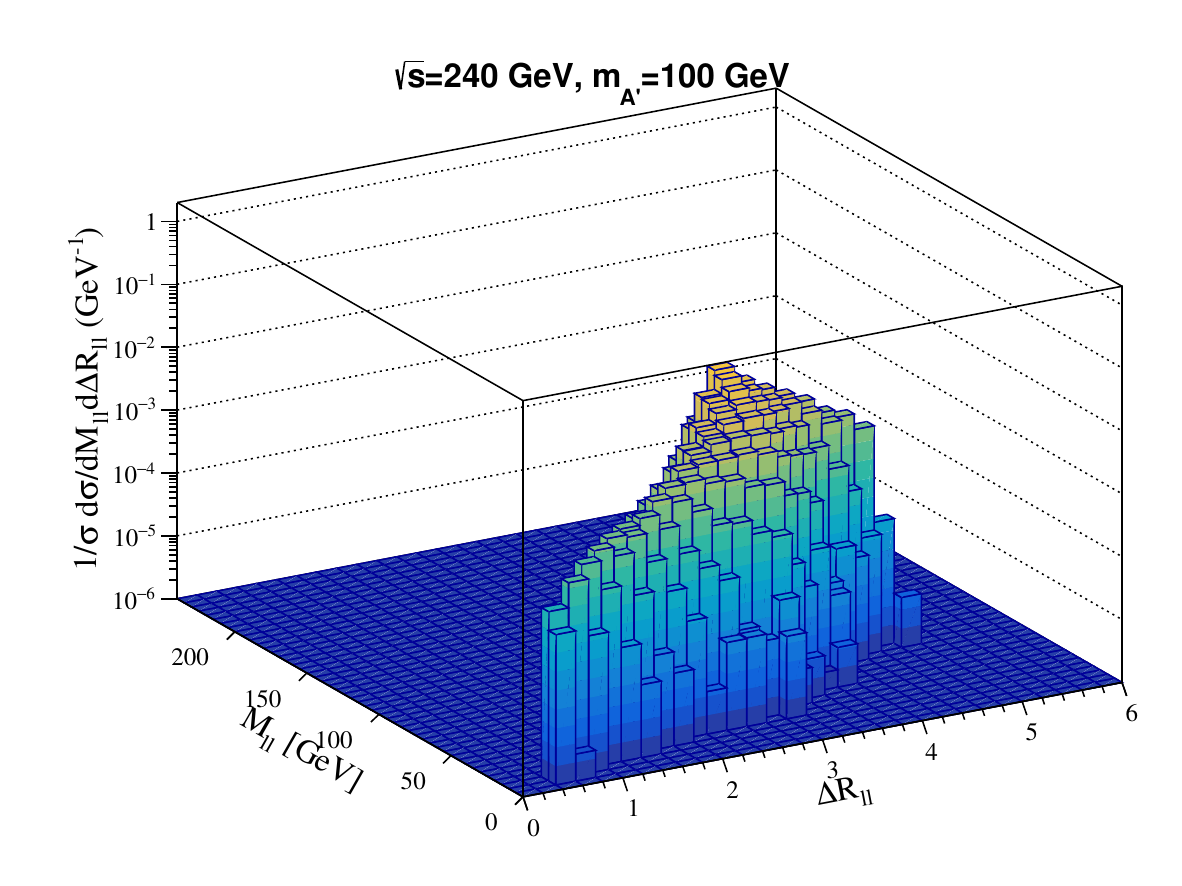}
        \caption{(c)}
    \end{subfigure}
    \hfill
    \begin{subfigure}{0.44\textwidth}
        \centering
        \includegraphics[width=\linewidth]{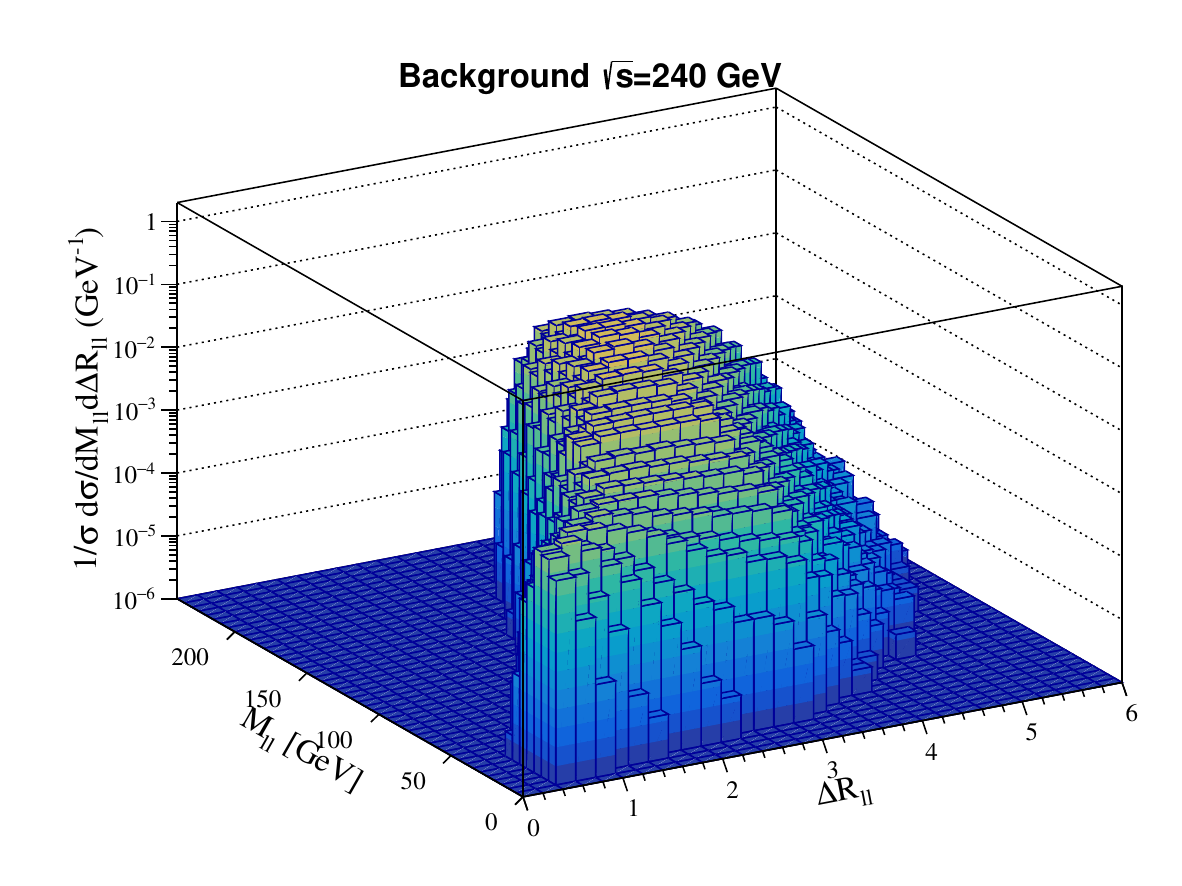}
        \caption{(d)}
    \end{subfigure}
\caption{Normalized distributions as functions of $M_{\ell\ell}$ and $\Delta R_{\ell\ell}$ for the signal and the dominant SM background process at the CEPC.}
    \label{fig:3b}
\end{figure}

We further investigate the normalized distributions as functions of $M_{\ell\ell}$ and $\Delta R_{\ell\ell}$, and present the corresponding three-dimensional distributions in Fig.~\ref{fig:3b}.
The signal process is shown for $M_{A'}=10~\mathrm{GeV}$ at $\sqrt{s}=91.2~\mathrm{GeV}$ and $M_{A'}=100~\mathrm{GeV}$ at $\sqrt{s}=240~\mathrm{GeV}$, together with the corresponding background processes.
As shown in Fig.~\ref{fig:3b}, at $\sqrt{s}=91.2~\mathrm{GeV}$, the signal events appear mainly around $\Delta R_{\ell\ell}\sim 3$ and $M_{\ell\ell}\sim 75~\mathrm{GeV}$, but the background distribution does not exhibit a similarly pronounced concentration.
At $\sqrt{s}=240~\mathrm{GeV}$, the signal is mainly concentrated in a relatively localized region of the $(M_{\ell\ell}, \Delta R_{\ell\ell})$ plane, while the background distribution covers a broader region.
The differences displayed in the three-dimensional plots can serve as a basis for the final event selection.

\section{\label{sec:result}Event Selection and Signal Significance}

\begin{figure}[H]
\centering
\begin{minipage}{0.45\textwidth}
    \centering
    \includegraphics[width=\linewidth]{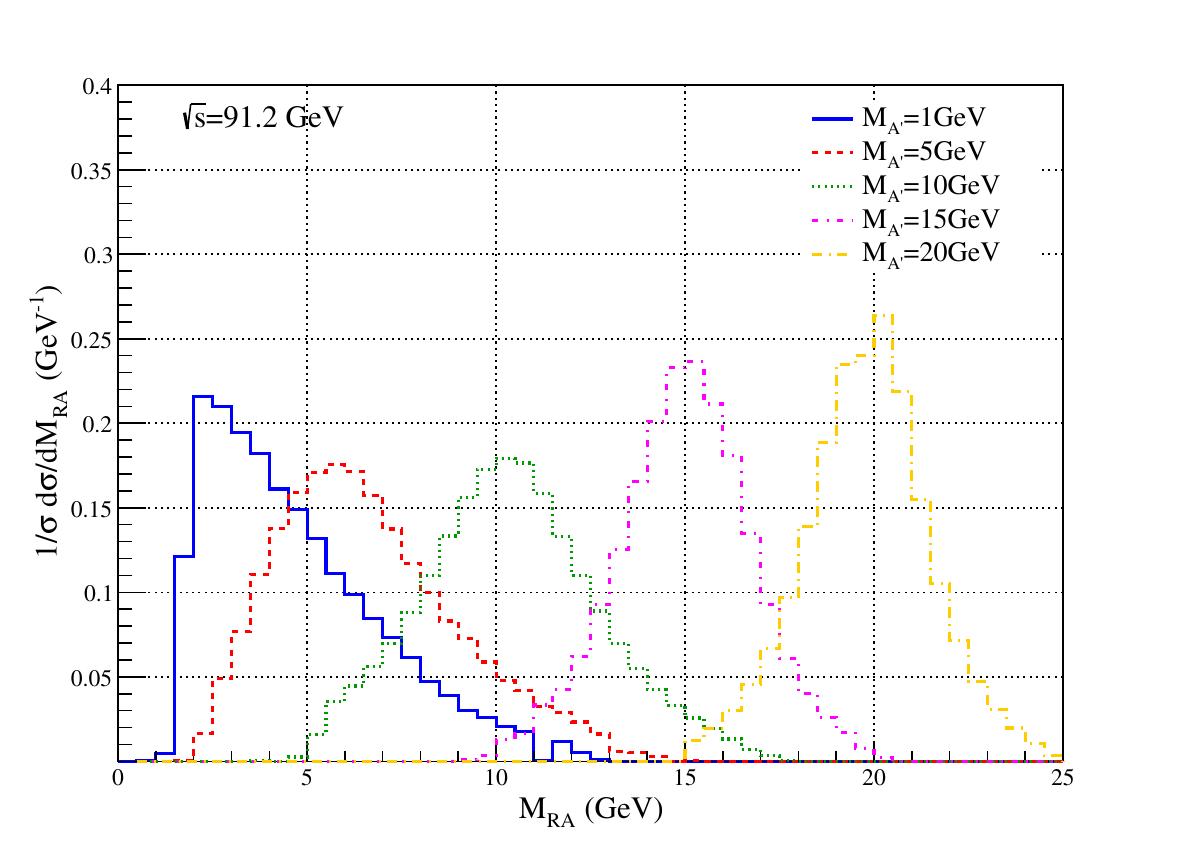}
    \caption*{(a)}
\end{minipage}
\hfill
\begin{minipage}{0.45\textwidth}
    \centering
    \includegraphics[width=\linewidth]{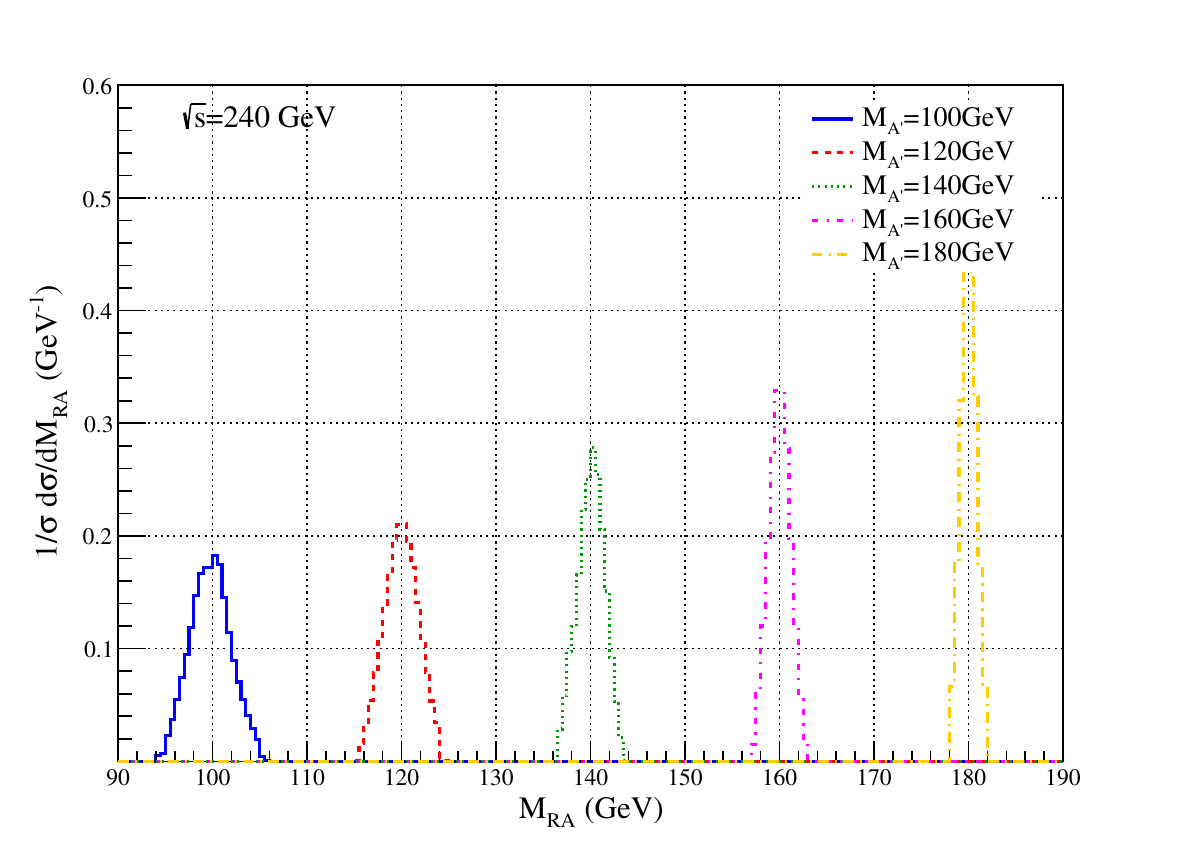}
    \caption*{(b)}
\end{minipage}
\caption{Normalized recoil-mass distributions for different dark photon masses in the process $e^+e^- \rightarrow \ell^+\ell^- A'$ for (a) $\sqrt{s}=91.2~\mathrm{GeV}$ and (b) $240~\mathrm{GeV}$.}
    \label{fig:zk}
\end{figure}

In addition to the kinematic distributions between the signal and SM background processes discussed in the previous section, we apply Gaussian smearing to the energies of the final state charged leptons to simulate the detector energy resolution using the following resolution function:
\begin{equation}
\frac{\delta(E)}{E}=\frac{A}{\sqrt{E/\mathrm{GeV}}}\oplus B .
\end{equation}
The detector resolution parameters are chosen with reference to the detector performance study in Ref.~\cite{Ai:2025cpj}.
For a relatively conservative estimate, we choose $A=3\%$ and $B=1\%$.
The dark photon mass is then reconstructed using the smeared four-momenta:
\begin{equation}
M_{\rm RA}=\sqrt{\left(p_{e^+}+p_{e^-}-p_{\ell^+}-p_{\ell^-}\right)^2}.
\end{equation}
The dark photon mass distributions reconstructed after detector resolution smearing are shown in Fig.~\ref{fig:zk}.
Owing to the detector resolution, the position of the signal peak may deviate from the corresponding dark photon mass, but this shift decreases gradually as the dark photon mass increases.
This feature will be taken into account when choosing the $M_{\mathrm{RA}}$ range for the subsequent significance analysis.

The statistical significance is calculated using the asymptotic expression~\cite{Cowan:2010js}
\begin{equation}
Z=\sqrt{2\left[(S+B)\ln\left(1+\frac{S}{B}\right)-S\right]},
\label{eq:significance}
\end{equation}
where $S$ and $B$ denote the expected numbers of signal and SM background events after applying all the selection criteria.
The number of signal events is calculated as
\begin{equation}
S=\sigma_{\mathrm{sig}}\mathrm{Br}(A'\rightarrow\chi\bar{\chi})\mathcal{L}_{\mathrm{int}},
\label{eq:signal_number}
\end{equation}
where $\sigma_{\mathrm{sig}}$ is the signal cross section after event selection, and $\mathcal{L}_{\mathrm{int}}$ is the designed integrated luminosity of the collider.
The integrated luminosity adopted in this analysis is based on the future collider projections discussed in Ref.~\cite{Ai:2025cpj}.

The decay branching ratios are calculated according to Eq.~\eqref{eq:br}, with $\alpha_D = 0.05$.
We choose $M_{A'}/m_\chi = 2.05$ to ensure that the invisible decay is allowed kinematically~\cite{Wang:2025cth}.
For the kinetic mixing parameter, we take $\epsilon = 5\times10^{-4}$, which is below the upper bounds on $\epsilon$ for the dark photon mass range considered here~\cite{Caputo:2026pdw}.
To examine the dependence on the dark matter mass, we also consider an alternative scenario with $M_{A'}/m_\chi = 3$, while keeping $\alpha_D=0.05$ and $\epsilon=5\times10^{-4}$ fixed.
For both dark-sector mass configurations considered in this work, the invisible decay channel dominates, with
\begin{equation}
\mathrm{Br}(A'\to\chi\bar{\chi}) > 99.99\%.
\end{equation}
Therefore, we conservatively approximate the invisible decay branching ratio as
\begin{equation}
\mathrm{Br}(A'\to\chi\bar{\chi}) = 0.9999.
\end{equation}

In addition to the basic selection criteria described in Sec.~\ref{sec:signal}, further kinematic requirements are imposed on both the signal and SM background processes.
For light dark photons, the recoil-mass windows are adjusted to account for the shift of the reconstructed signal peak from $M_{A'}$.
For heavier dark photons, the recoil-mass windows are centered on $M_{A'}$.
Then, using the three-dimensional distribution shown in Fig.~\ref{fig:3b}, an initial selection on $M_{\ell\ell}$ and $\Delta R_{\ell\ell}$ is made for each dark photon mass point and further refined on this basis.
The cross sections for both the signal and background processes are obtained under the same selection conditions.
The selection criteria for each mass point are summarized in Table~\ref{tab:cut_mass}, together with the resulting signal and SM background event yields and the significance.

\begin{table}[H]
\centering
\caption{Expected event yields and statistical significances for the signal and background at $\sqrt{s}=91.2~\mathrm{GeV}$ and $\mathcal{L}_{\mathrm{int}}=100~\mathrm{ab}^{-1}$.}
\label{tab:cut_mass}
\begin{tabular}{ccccccc}
\toprule
$M_{A'}$ (GeV) &
$M_{\rm RA}$ range &
$\Delta R_{\ell\ell}$ &
$M_{\ell\ell}$ (GeV) &
$S$ &
$B$ &
$Z$ \\
\midrule
1  & $0.9-1.7$  & $3-5$     & $70-90$ & 225.2 & 9.7    & 32.1 \\
2  & $1.6-2.8$  & $3-5$     & $69-90$ & 290.7 & 31.9   & 30.0 \\
3  & $2.1-3.9$  & $3-5$     & $69-90$ & 385.2 & 74.7   & 29.8 \\
4  & $2.1-5.9$  & $3-5$     & $69-89$ & 555.3 & 245.6  & 27.7 \\
5  & $3-7$      & $2.8-5$   & $68-88$ & 534.5 & 428.3  & 22.0 \\
6  & $4.2-8.2$  & $2.8-5$   & $66-86$ & 475.3 & 628.5  & 17.0 \\
7  & $4.8-9.2$  & $2.8-5$   & $63-85$ & 468.6 & 944.1  & 14.1 \\
8  & $6-8.3$    & $2.7-5$   & $62-82$ & 399.1 & 1149.2 & 11.1 \\
9  & $6.8-11.2$ & $2.7-4.8$ & $61-84$ & 365.7 & 1377.9 & 9.4 \\
10 & $7.8-12.2$ & $2.6-4.8$ & $58-82$ & 341.1 & 1557.0 & 8.3 \\
11 & $8.6-13.4$ & $2.6-4.7$ & $51-82$ & 344.3 & 2596.3 & 6.6 \\
12 & $9.3-14.7$ & $2.6-4.2$ & $53-81$ & 310.2 & 2959.3 & 5.6 \\
13 & $9.7-16.3$ & $2.6-4.2$ & $48-79$ & 311.9 & 4284.3 & 4.7 \\
14 & $11.7-16.3$ & $2.6-4.2$ & $46-78$ & 240.7 & 3254.3 & 4.1 \\
15 & $12.9-17.1$ & $2.6-4.0$ & $46-76$ & 199.2 & 3185.5 & 3.5 \\
16 & $13.9-18.1$ & $2.6-4.0$ & $45-76$ & 180.2 & 3349.4 & 3.1 \\
17 & $14.9-19.1$ & $2.6-4.0$ & $44-76$ & 162.8 & 3542.4 & 2.7 \\
18 & $15.9-20.1$ & $2.6-4.0$ & $42-75$ & 147.2 & 3758.7 & 2.4 \\
19 & $16.9-21.1$ & $2.6-4.0$ & $42-74$ & 131.7 & 3840.0 & 2.1 \\
20 & $17.9-22.1$ & $2.6-4.0$ & $40-72$ & 118.6 & 3988.9 & 1.8 \\
\bottomrule
\end{tabular}
\end{table}

Table~\ref{tab:cut_mass} presents the significance for each mass point in the studied range at $\sqrt{s}=91.2~\mathrm{GeV}$. 
The significance decreases with increasing $M_{A'}$ due to the decrease in the total signal cross section.
It remains above $3\sigma$ for $M_{A'} \le 16~\mathrm{GeV}$ and reaches $1.8\sigma$ at $M_{A'}=20~\mathrm{GeV}$.
For the $\sqrt{s}=240~\mathrm{GeV}$ configuration, taking $M_{A'}=100~\mathrm{GeV}$ as an example, we apply the detector resolution parameters described above, together with the angular separation requirement $1<\Delta R_{\ell\ell}<3$, the invariant-mass cut $M_{\ell\ell}<140~\mathrm{GeV}$, and the $M_{RA}$ range $98~\mathrm{GeV}<M_{\rm RA}<102~\mathrm{GeV}$. The invisible decay branching ratio is included in the calculation of the expected signal yield.
We obtain a signal cross section of $\sigma_{\mathrm{sig}}=1.37\times10^{-10}~\mathrm{pb}$ and a background cross section of $\sigma_B=1.04\times10^{-2}~\mathrm{pb}$. 
At the design luminosity of $\mathcal{L}_{\mathrm{int}}=6~\mathrm{ab}^{-1}$, these cross sections correspond to approximately $8.22\times10^{-4}$ signal events and $6.24\times10^{4}$ background events.
Thus, the expected signal yield is far below one event at the designed luminosity.
Obtaining even a single signal event would require an integrated luminosity of approximately $7.30\times10^{3}~\mathrm{ab}^{-1}$, which is far beyond the projected CEPC luminosity at this energy. 
These results indicate that the dilepton recoil channel has very limited sensitivity to heavier dark photons at $\sqrt{s}=240~\mathrm{GeV}$ under the benchmark scenario considered here.

\begin{figure}[H]
    \centering
    \begin{subfigure}[b]{0.45\textwidth}
        \centering
        \includegraphics[width=\textwidth]{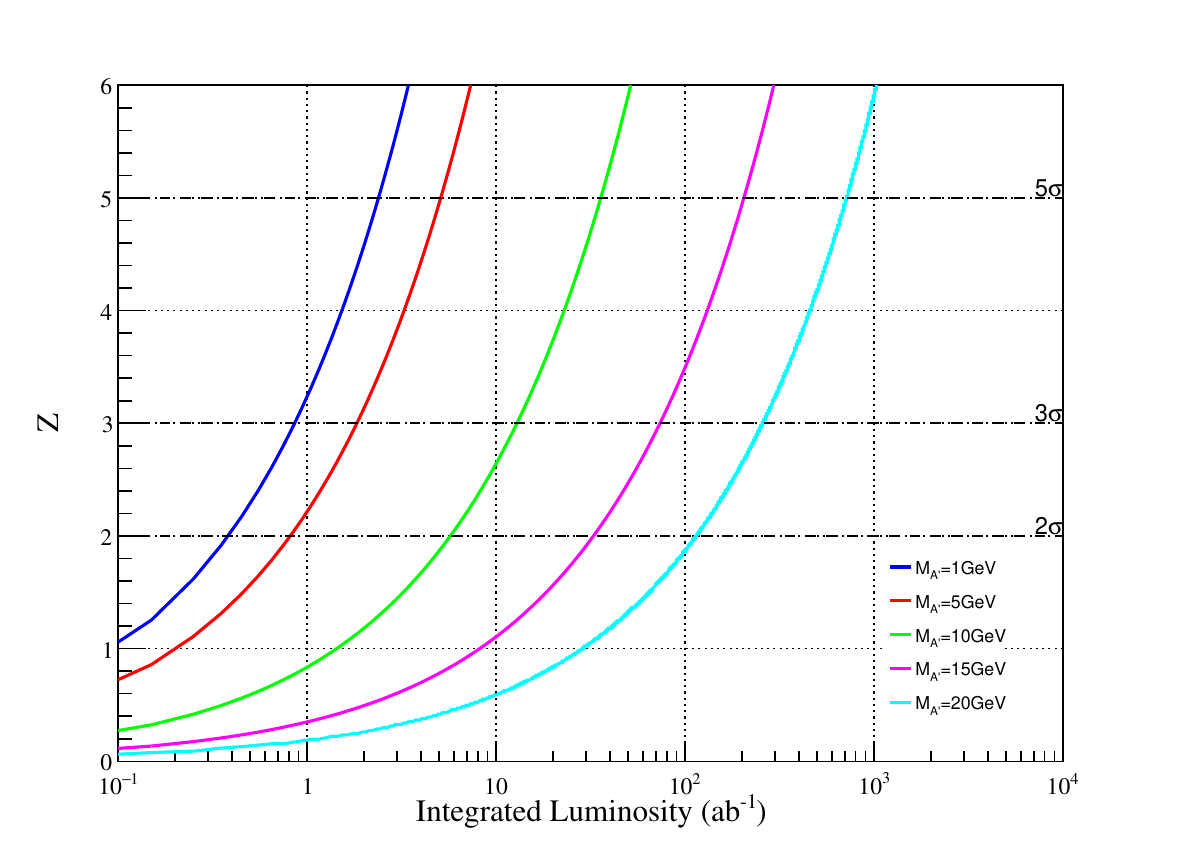}
        \caption{(a)}
        \label{fig:zq}
    \end{subfigure}
    \hfill
    \begin{subfigure}[b]{0.40\textwidth}
        \centering
        \includegraphics[width=\textwidth]{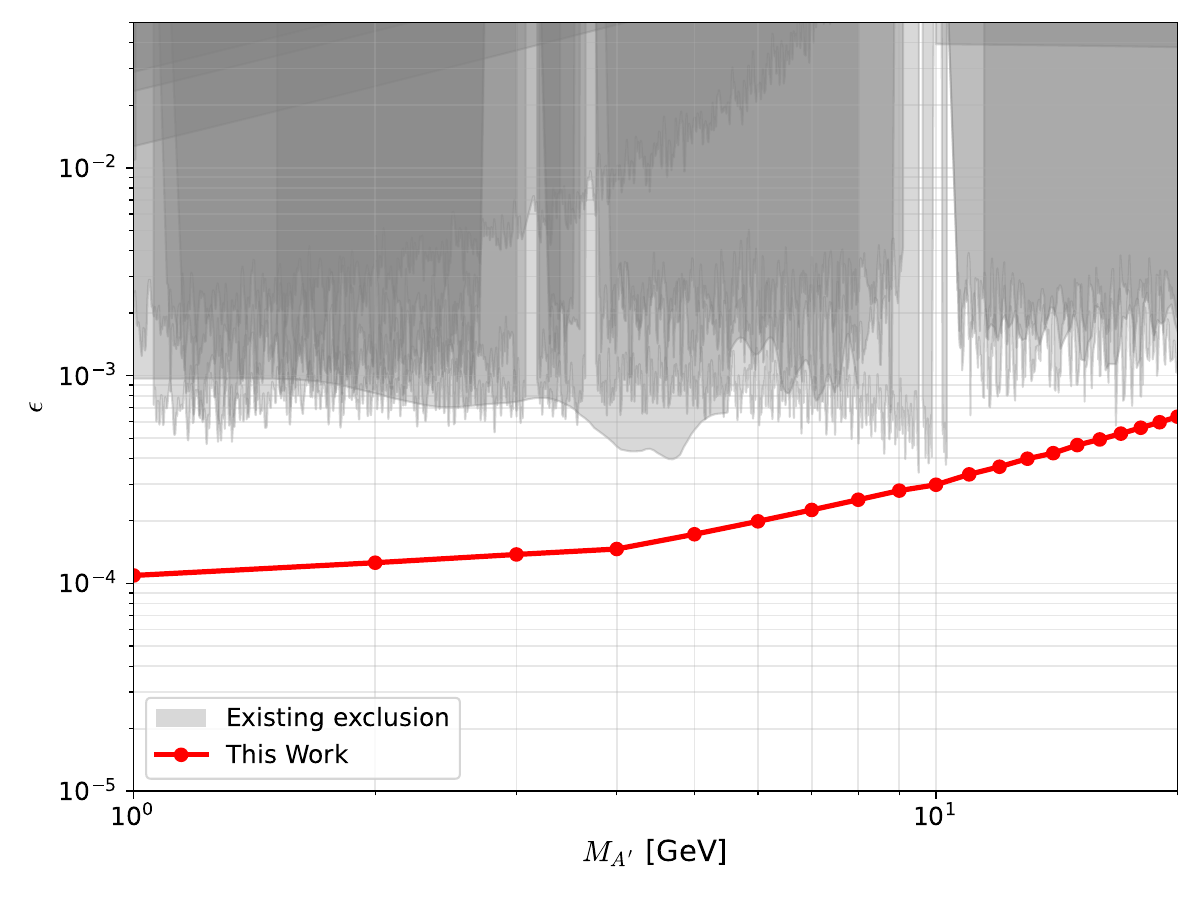}
        \caption{(b)}
        \label{fig:epsilon}
    \end{subfigure}

\caption{(a) Statistical significance as a function of the integrated luminosity; (b) Shaded regions: existing constraints obtained using the tool~\cite{Ilten:2018crw}. Red points: minimum required $\epsilon$ values to achieve $3\sigma$ significance in this work.}
    \label{fig:ss}
\end{figure}

In contrast, the $\sqrt{s}=91.2~\mathrm{GeV}$ configuration yields better results.
Fig.~\ref{fig:ss} (a) shows the variation of $Z$ with $\mathcal{L}_{\mathrm{int}}$ for $M_{A'}=1, 5, 10, 15$, and $20~\mathrm{GeV}$.  
The minimum integrated luminosities required to reach $3\sigma$ are approximately $0.87$, $1.86$, $13.1$, $73.5$, and $2.78\times10^{2}~\mathrm{ab}^{-1}$ for $M_{A'}=1$, $5$, $10$, $15$, and $20~\mathrm{GeV}$, respectively.
The required integrated luminosity increases rapidly with $M_{A'}$, which indicates a pronounced decrease in sensitivity at larger dark photon masses. 
In particular, for $M_{A'}\leq16~\mathrm{GeV}$, a $3\sigma$ significance can be reached with an integrated luminosity below $100~\mathrm{ab}^{-1}$, whereas $M_{A'}=20~\mathrm{GeV}$ requires more than $100~\mathrm{ab}^{-1}$ and thus cannot reach this significance at the benchmark luminosity.
This is consistent with the significance given in Table~\ref{tab:cut_mass}.
Fig.~\ref{fig:ss}(b) shows the minimum required $\epsilon$ values to achieve $3\sigma$ significance for dark photon masses in the range $1~\mathrm{GeV} \le M_{A'} \le 20~\mathrm{GeV}$, together with the existing constraints.
As can be seen, in this mass range, the minimum required $\epsilon$ values calculated in this work are not excluded by the existing constraints.

\section{\label{sec:conclusion}Summary}

The dark photon is a vector-portal candidate for physics beyond the SM and may mediate interactions between the SM and the dark sector. 
In this paper, we investigate the production of an invisible dark photon via the process $e^+e^- \to \ell^+\ell^- A'$ ($\ell=e,\mu$) at the CEPC, with the invisible decay $A'\to\chi\bar{\chi}$.
We calculate the signal cross sections as functions of the dark photon mass and the center-of-mass energy.
We then compare the kinematic distributions of the signal and background processes for several kinematic variables, including $M_{\ell\ell}$, $p_T^{\ell\ell}$, $\cos\theta_{\ell\ell}$, and $\Delta R_{\ell\ell}$, and analyze the differences between them. 
The cross section as a function of the dilepton invariant mass $M_{\ell\ell}$ and the dilepton separation $\Delta R_{\ell\ell}$ is further calculated, and the corresponding distributions are presented.
The resulting distributions exhibit a clear separation between the signal and the background and thus provide additional discriminating power.
To account for the detector resolution at the collider, we apply smearing to the energies of the final-state charged leptons and reconstruct the dark photon mass distributions.
The selection cuts are optimized according to the kinematic distributions between the signal and the background, and the significance is calculated for each dark photon mass point after including the invisible decay branching ratio.
At $\sqrt{s}=91.2~\mathrm{GeV}$ with an integrated luminosity of $100~\mathrm{ab}^{-1}$, the significance exceeds $3\sigma$ for dark photon masses in the range $1~\mathrm{GeV} \le M_{A'} \le 16~\mathrm{GeV}$, indicating that this process has the potential to discover dark photons. 
At $\sqrt{s}=240~\mathrm{GeV}$, for dark photon masses in the range $20$--$200~\mathrm{GeV}$, the expected number of events after applying the selection cuts is less than one, making it impossible to probe dark photons through this process at the collider.

\section*{Acknowledgements}
This work was supported in part by the National Natural Science Foundation of China (Grants No. 12105162, 12305106), the Natural Science Foundation of Shandong Province (Grants No. ZR2021QA058, ZR2021QA040) and the Youth Innovation Technology Project of Higher School in Shandong Province (2023KJ146).

\end{document}